\documentclass{acm_proc_article-sp}

\usepackage{balance}  % to better equalize the last page
\usepackage{graphics} % for EPS, load graphicx instead
\usepackage{amsmath}
\usepackage{amssymb}
\usepackage{subfigure}
\usepackage{multirow}
\usepackage{cite}
\usepackage{pbox}

\usepackage{makecell}

\permission{Copyright is held by the International World Wide Web Conference Committee (IW3C2). IW3C2 reserves the right to provide a hyperlink to the author's site if the Material is used in electronic media.}
\conferenceinfo{WWW 2016,}{April 11--15, 2016, Montr\'eal, Qu\'ebec, Canada.} 
\copyrightetc{ACM \the\acmcopyr}
\crdata{978-1-4503-4143-1/16/04. \\
DOI: http://dx.doi.org/10.1145/2872427.2883063l}

\begin{document}

\title{Social Networks Under Stress}

\numberofauthors{3} 

\author{
% 1st. author
\alignauthor
Daniel M. Romero\\
       \affaddr{University of Michigan}\\
       \affaddr{Ann Arbor, MI}\\
       \email{drom@umich.edu}
% 2nd. author
\alignauthor
Brian Uzzi\\
       \affaddr{Northwestern University and Northwestern University Institute on Complex Systems (NICO)}\\
        \affaddr{Evanston, IL}\\
       \email{uzzi@northwestern.edu }
% 3rd. author
\alignauthor Jon Kleinberg\\
       \affaddr{Cornell University}\\
        \affaddr{Ithaca, NY}\\
       \email{kleinber@cs.cornell.edu }
}

\maketitle
\begin{abstract}
Social network research has begun to take advantage of fine-grained communications regarding coordination, decision-making, and knowledge sharing.  These studies, however, have not generally analyzed how external events are associated with a social network's structure and communicative properties.  Here, we study how external events are associated with a network's change in structure and communications.  Analyzing a complete dataset of millions of instant messages among the decision-makers in a large hedge fund and their network of outside contacts, we investigate the link between price shocks, network structure, and change in the affect and cognition of decision-makers embedded in the network.  When price shocks occur the communication network tends not to display structural changes associated with adaptiveness.  Rather, the network ``turtles up".  It displays a propensity for higher clustering, strong tie interaction, and an intensification of insider vs. outsider communication.  Further, we find changes in network structure predict shifts in cognitive and affective processes, execution of new transactions, and local optimality of transactions better than prices, revealing the important predictive relationship between network structure and collective behavior within a social network.

\end{abstract}

\category{H.2.8}{Database Management}{Database applications---Data mining}

\keywords{Social Networks, Organizations, Temporal Dynamics, Collective Behavior.} % NOT required for Proceedings

\section{Introduction}

The emergence of detailed, time-resolved data on large social networks makes it possible to study their structure and dynamics in new ways.  Work has identified stable network structures, communities, influentials in networks, information flows, networked teams \cite{Newman-sirev,Jackson-book,Lazer-science,Easley-book,Newman-book}, and the temporal evolution of large social networks \cite{Barabasi-book,Chakrabarti-Faloutsos,Kossinets_2006, Kossinets_2008, Leskovec, Saavedra, Qing, Ahn, Peel, Weng, Viswanath}.

Despite the richness of the available datasets and the computational techniques for analyzing them, work has generally not focused on a crucial feature of dynamic behavior --- how changes in the social network's structure are linked to external shocks.  Little research thus far has examined how social networks operate in a reactive capacity, as they respond to such stimuli from their broader environments.  By external shocks we mean events that are extreme relative to average events, or unexpected \cite{Gilbert}.  Such questions are critical to understanding a social network's capacity to respond to uncertainty when the membership of the network remains essentially fixed.  The real-life conditions reflecting these dynamics are varied, including the response of intelligence and law enforcement personnel to emergency situations such as terrorist attacks \cite{Butts}, health professionals responding to outbreaks \cite{Dutta}, or organizations facing sudden competitive threats or ``normal accidents" \cite{Perrow}.  The effect has been to leave a set of basic questions largely unanswered.  How does a network respond to external shocks?  What can shock-related responses reveal about adaptive collective behavior?

Current frameworks suggest different conjectures for relating social network structure to external risks.  Networks could open up structurally, with actors tapping acquaintances who are most likely to provide novel information and diverse perspectives on problem-solving \cite{Burt92}.  Experiments show that persons facing threatening job changes, for example, disproportionately turn to their weak ties and reduced their triadic closure \cite{Menon, Smith}, broadening their options and access to novel job information \cite{Granovetter}.  Conversely, external shocks could be associated with networks ``turtling up," with actors relying on well-worn sources of information, network insiders, and highly clustered relationships \cite{Coleman, Granovetter85} that promote trustworthiness but narrow social cognition \cite{Menon,Ellis, Coleman, Granovetter}. Further, little is known about how individual cognitive and affective reactions aggregate in a network and whether collective reactions are better predicted by knowledge of the shock or structural changes in the network \cite{Butts}.   

We address these questions in the context of an organization's social network \cite{Adamic,Brass, Diesner, Dodds,Kilduff,Klimt,Uzzi}.  We consider the complete instant-messaging corpus, including content, among the decision-makers of a hedge fund and their outside contacts.  The analysis of financial investment behavior is of interest because of its criticality to the economy and society \cite{Creevy, Lo, Preis, Saavedra13, Mann, Fama}. 

We focus on a few central questions.  First, we study the link between environmental changes --- encoded in the movement of the prices of specific stocks at particular points in time --- and changes in the network structure, specifically the subgraph consisting of instant messages about these particular stocks during the times in question. For financial services firms, price shocks can represent threatening and stressful external stimuli. Experiments measuring the stress level of traders using conductance technology show that price changes lead to involuntary and significant physiological changes indicative of stress.  Shifts in prices cause traders to have increased heart rates and electrodermal responses relative to control conditions \cite{Lo}.   Qualitative observations of traders echo experimental results.  In the industry, the VIX, a widely used measure of market price volatility, is referred to as the ``Fear Index \cite{Whaley}."  

Following sociological theory, one expectation for the association between price shocks and network change is that price shocks are associated with a propensity to activate connections that improve access to novel information and manage risk.  In this case, it would be expected that actors in the network would rely relatively more on weak ties, network outsiders, and relationships with low closure.  Conversely, shocks may be associated with reflective network structure and behavior \cite{Staw}.  Following this view, expectations are that actors disproportionately favor contacts repeatedly seen in the past, in-group rather than out-group relations, and highly interconnected contacts.   In the face of shocks, our analysis supports the conclusion that networks ``turtle up" structurally, exhibiting high levels of clustering, out-group communication, and strong ties.

Where our first question explores the link between external factors and
network structure, our second main question asks whether these changes
to the network structure can yield additional insight into 
the behavior of the organization --- beyond what is provided
by the external changes themselves.
Specifically, if we seek to predict how the organization will respond
to a price change, does knowing properties of the network structure improve the
performance of the prediction even if we already have access to the
time series of stock prices?

We show that knowledge of the network provides 
strong improvements in prediction
of collective behavior for both individual-level actions and firm-level actions.
We evaluate individual-level actions by analyzing the way traders express themselves in the collection of
instant messages, using a standard set of linguistic measures.
These measures show that when a stock's price changes significantly
(in either the positive or negative direction), the messages associated
with that stock display increased emotion and cognitive complexity.
While these changes can already be inferred to a limited extent from
the price changes alone, we are able to predict their direction and
magnitude much more effectively when we incorporate network-level features
into the prediction.
We find analogous effects in the context of
firm-level actions, where we analyze the trading decision employees make for the firm. We introduce a simple measure of local optimality
in the trading price and show
that network features provide significantly
improved performance in predicting whether the firm will perform a
locally optimal action. We find a similar pattern when we attempt to predict whether trades that have not been observed for a number of days are suddenly traded. 

Taken together, our findings suggest a key role for electronic communication networks in the analysis of external shocks and events affecting an organization:
the network displays a consistent set of changes in response to 
these external events, emphasizing strong ties and clustered structures;
and knowledge of the network structure provides significant leverage
in predicting the organization's collective behavior.

\section{Data}

Our data include the complete instant-messaging communication history among the personnel at a hedge fund, 2010-2011. The data consist of approximately 22 million instant messages (IMs) sent by 8,646 people, of whom 184 are employees of the hedge fund, and the rest are outside contacts.  We use the full content of each message and unique person and time of transmission identifiers to analyze the IMs.  We know each trade completed at the firm, the stock symbol involved, and date and time of execution.  All transactions in our data were executed by the fund's employees, not algorithms.  On average, 559 transactions were performed daily.  We merged the above data with public data on daily stock prices. All of our data analysis was consistent with guidelines from the relevant Institutional Review Board

\section{Measures}

\subsection{Network}

IMs define a network of information exchange where nodes are communicating actors and IMs between actors define edges. Our primary interest is in the subgraphs of the larger network designated by mentions of particular stocks at particular points in time.  For a stock symbol $s$ and a day $d$, we define an undirected graph $G_{s,d}$ as follows: for each IM between company insiders that mentions stock symbol $s$ on day $d$, we include the two participants of the IM as nodes in $G_{s,d}$, and we join them by an edge. We eliminate parallel edges, so that two nodes in $G_{s,d}$ will have at most one edge between them. Multiple messages between pairs were examined when relevant.  $G_{s,d}$ thus consists of all employees who mentioned stock $s$ on day $d$, with messages containing $s$ on day $d$ forming the edges between these people.
\begin{table*}
\centering
\caption{Results form OLS regressions of the form $f(G_{s,d}) = \beta_0|\Delta_{s,d}| + \beta_1f(G_{s,d-1}) +   \beta_2f(G_{s,d-2}) +   \beta_3VIX + r_{s,d}$. Regressions include fixed effects at the stock and day of the week level. Each column represents a regression with $f$ as the dependent variable and the rows show the value and significance of the independent variable. Asterisks indicate coefficient significance (***:p-value $< 0.0001$).}
\label{regressions_fe_stock}
\begin{tabular}{c|c|c|c|c}
\Xhline{4\arrayrulewidth}
& \bf{Model} 1 & \bf{Model} 2 & \bf{Model 3} & \bf{Model 4}\\
 \bf{Independent Variables} &  \bf{$f = $Nodes} & \bf{ $f = $Clustering} & \bf{$f = $Perc. border edges}  & \bf{$f = $Strength of ties} \\
\Xhline{4\arrayrulewidth}
 Stock price change &  $0.0540^{***}$ &  $0.1590^{***}$&$-0.0009^{***}$ & $0.0010^{***}$  \\
 \hline
  $f$-lag(-1)  & $-0.0009^{***}$ & $0.0089^{***}$ & $0.1415^{***}$& $0.0448^{***}$ \\
 \hline
  $f$-lag(-2)& $0.0644^{***}$& $0.0026$ & $0.0959^{***}$&  $0.0270^{***}$\\
   \hline
 VIX  & $-0.0034^{***}$ & $-0.0213^{***}$ &$-0.0006^{***}$ & $-0.0001$ \\
 \hline
  5 Day of week fixed effects  & Y & Y &Y & Y \\
 \hline
   Stock fixed effects & Y & Y &Y & Y \\
 \hline
\end{tabular}
\end{table*}

\begin{figure}
\centering
\includegraphics[width =.5 \textwidth]{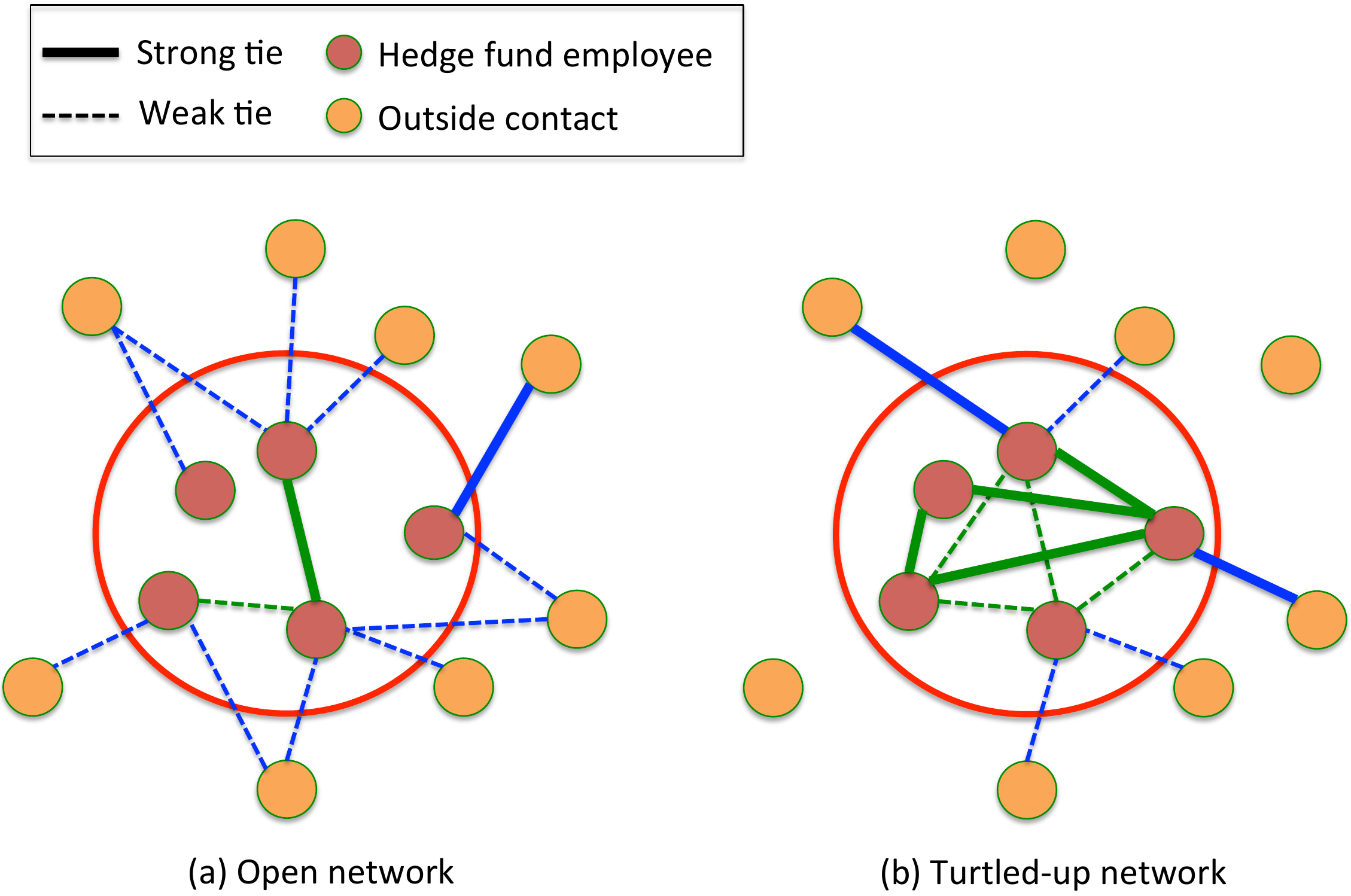} 
\caption{Structure of \emph{open} (a) and \emph{turtled up} (b) networks.  Turtled up networks exhibit a propensity towards strong ties over weak ties, high clustering, and insider vs. outsider  links.  Networks tend to turtle up on days with large or unexpected stock price changes. \label{network_illustration}}
\end{figure}

We define an edge to be \emph{internal} if both nodes are company insiders and \emph{border} if one node is a company outsider.  All IMs include one company employee because communication between company outsiders cannot be recorded. Our definition of $G_{s,d}$ uses only internal edges; the analogous construction using both internal and border edges results in a larger graph $G_{s,d}^+$ that contains $G_{s,d}$.

The notation and terminology used in the analysis for graph size and connectivity and quantifications of  \emph{turtled up} or \emph{open} networks is as follows. Let $N_{s,d}$ and $E_{s,d}$ be the number of nodes and edges respectively in $G_{s,d}$. We normalize relevant measures in relation to comparable quantities in the data; in particular, for a function $f(G_{s,d})$ that we compute on $G_{s,d}$, we will define $\nu(f(G_{s,d}))$ to be the ratio of $f(G_{s,d})$ to the average value of $f(G_{s',d'})$ for all pairs $(s',d')$ such that $N_{s',d'} = N_{s,d}$ and $d' < d$. This lets us discuss whether $f(G_{s,d})$ is large or small relative to other graphs from previous days with the same number of nodes.  We can define the analogous normalization based on the number of edges; we write $\epsilon(f(G_{s,d}))$ to denote the ratio of $f(G_{s,d})$ to the average value of $f(G_{s',d'})$ for all pairs $(s',d')$ such that $E_{s',d'} = E_{s,d}$ and $d' < d$.  Furthermore, since some stocks tend to be more popular in traders' conversations than others, we normalize the number of nodes.  We define $\bar{N}_{s,d}$ as the ratio of $N_{s,d}$ to the average values of $N_{s,d'}$ for all $d' < d$.

\begin{table*}
\centering
\caption{Results form OLS regressions of the form $f(G_{s,d}) = \beta_0|\Delta_{s,d}| + \beta_1f(G_{s,d-1}) +  \beta_2f(G_{s,d-2}) +  \beta_3VIX  + r_{s,d}$. Regressions include fixed effects at the industry and day of the week level. Each column represents a regression with $f$ as the dependent variable and the rows show the value and significance of the independent variable. Asterisks indicate coefficient significance (***:p-value $< 0.0001$).}
\label{regressions_fe_industry}
\begin{tabular}{c|c|c|c|c}
\Xhline{4\arrayrulewidth}
& \bf{Model} 1 & \bf{Model} 2 & \bf{Model 3} & \bf{Model 4}\\
 \bf{Independent Variables} &  \bf{$f = $Nodes} & \bf{$f = $Clustering} & \bf{$f = $Perc. border edges}  & \bf{$f = $Strength of ties} \\
\Xhline{4\arrayrulewidth}
 Stock price change &  $0.0557^{***}$ &  $0.1925^{***}$&$-0.0004^{***}$ & $0.0009^{***}$  \\
 \hline
  $f$-lag(-1)  & $0.2579^{***}$ & $0.0209^{***}$ & $0.2008^{***}$& $0.0729^{***}$ \\
 \hline
  $f$-lag(-2)& $0.1277^{***}$& $0.0146^{***}$ & $0.1550^{***}$&  $0.0542^{***}$\\
   \hline
 VIX  & $-0.0038^{***}$ & $-0.0230^{***}$ &$-0.0006^{***}$ & $-0.0001$ \\
 \hline
   5 Day of week fixed effects  & Y & Y &Y & Y \\
 \hline
   Industry fixed effects & Y & Y &Y & Y \\
 \hline
\end{tabular}
\end{table*}

Beyond the number of nodes and edges, a basic quantity is connectivity --- whether $G_{s,d}$ is connected or whether most nodes belong to few connected components.  We write $L_{s,d}$ for the fraction of nodes in the largest connected component of $G_{s,d}$, and $K_{s,d}$ for the minimum number of components required to account for at least 90\% of nodes in $G_{s,d}$.

Sociological theory distinguishes ``closed" network structures as containing many triangles of relationships (friends of friends link to each other) and a high proportion of strong ties (high-frequency relationships).  We are referring to networks with this structure as \emph{turtled up} networks. Open network structures contain few triangles of relationships and have relatively high numbers of weak ties.  
Figure \ref{network_illustration} illustrates key differences between closed or turtled up networks and open networks.  

A measure for capturing triangles of relationship is the {\em clustering coefficient}; formally the fraction of pairs of a node's neighbors that are connected by an edge.  We define $C_{s,d}$ as the average clustering coefficient over all nodes in $G_{s,d}$.  To capture strength of ties we use frequency of communication, which is a relevant measure for the structure of a social network \cite{DeChoudhury}. For each node $x$ in $G_{s,d}$, we consider the set of all nodes $y$ with whom $x$ has participated in any messages on days $d' < d$, and we sort these nodes $y$ in descending order by the number of such messages they have participated in $x$.  We define $U_{x,d,\alpha}$ as the highest $\alpha$ fraction of this sorted list: the $\alpha$ fraction of $x$'s communication partners from the prior day $d$, measured by communication volume these are $x$'s strongest edges.  We quantify whether $G_{s,d}$ favors strong or weak ties using the measure $S_{s,d,\alpha}$, the fraction of edges $(x,y)$ for which $y \in U_{x,d,\alpha}$.  To the extent that closed and open structures are related to an actor's reach for information across boundaries \cite{Burt92}, we define the {\em openness} $O_{s,d}$ as the fraction of edges in $G_{s,d}^+$ that are border edges.

\subsection{Shocks}

To define the extremeness and unexpectedness of price shocks, we defined for each stock $s$ and day $d$, $a_{s,d}$ and $b_{s,d}$ to be the opening and closing prices respectively of stock $s$ on day $d$.  We define $\displaystyle{\Delta_{s,d} = \frac{b_{s,d} - a_{s,d}}{a_{s,d}}}$  as the proportional change in the price of $s$ on day $d$.  

Some price changes are disruptive --- the price change magnitudes are greater than recent changes, and therefore they can be characterized as unexpected. They intuitively correspond to shocks \cite{Gilbert}. We operationalize the notion of a ``shock'' as follows: a stock-day pair $(s,d)$ is an \emph{$x$-shock} if $|\Delta_{s,d}| > x$ and $|\Delta_{s,d'}| \leq x$ for $d' = d-3, d-2, d-1$. That is, $(s,d)$ is an $x$-shock if stock $s$'s price change on day $d$ was higher than $x$, and its price change was lower than $x$ on the previous three days. We investigate how continuous values of $\Delta_{s,d}$ and discrete $x$-shocks relate to the properties of $G_{d,s}$.

\section{Findings}

\subsection{Price Changes and Network Dynamics}

\begin{figure*}
\centering

\subfigure[Average clustering ($\nu(C_{s,d})$) ]{
\includegraphics[width =.31\textwidth]{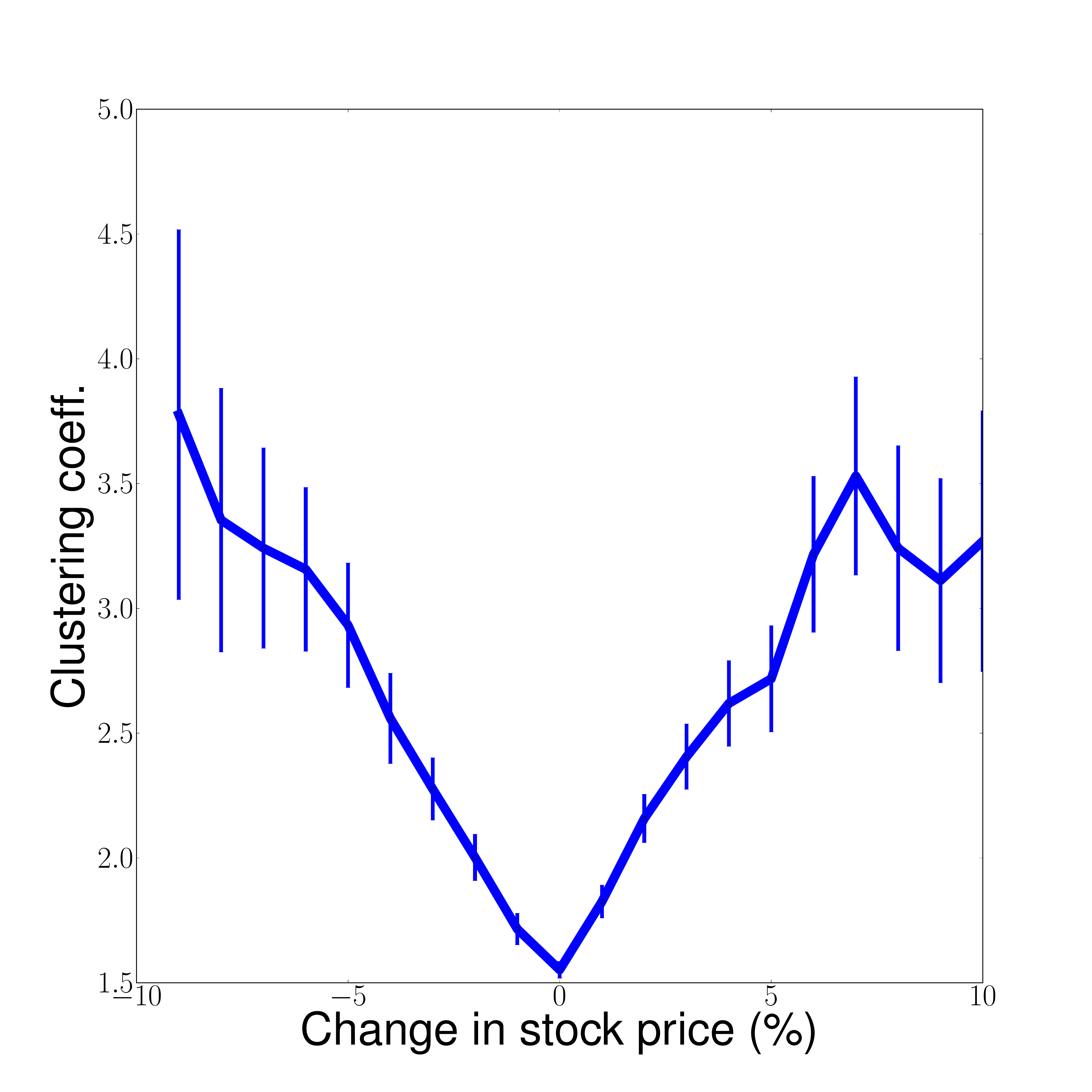} 
\label{clusteringCoeff_cNodes_InternalEdges_2}
}
\subfigure[Strength of ties ($S_{s,d,.1}$)]{
\includegraphics[width =.31 \textwidth]{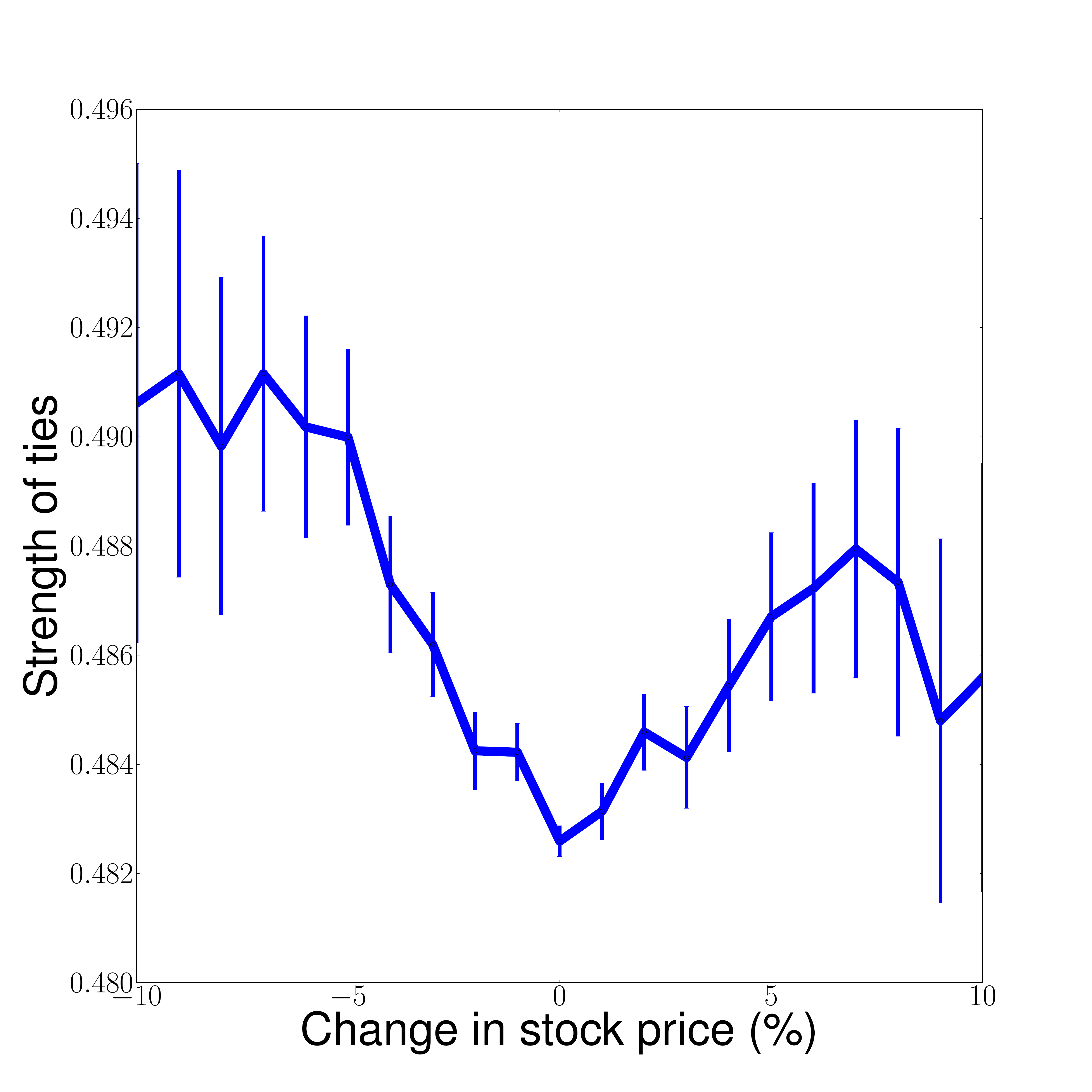} 
\label{Strength_of_ties_2_FracTopConnection_aggregate_ComputeEachUser}
}
\subfigure[Percentage of border edges ($O_{s,d}$)]{
\includegraphics[width =.31 \textwidth]{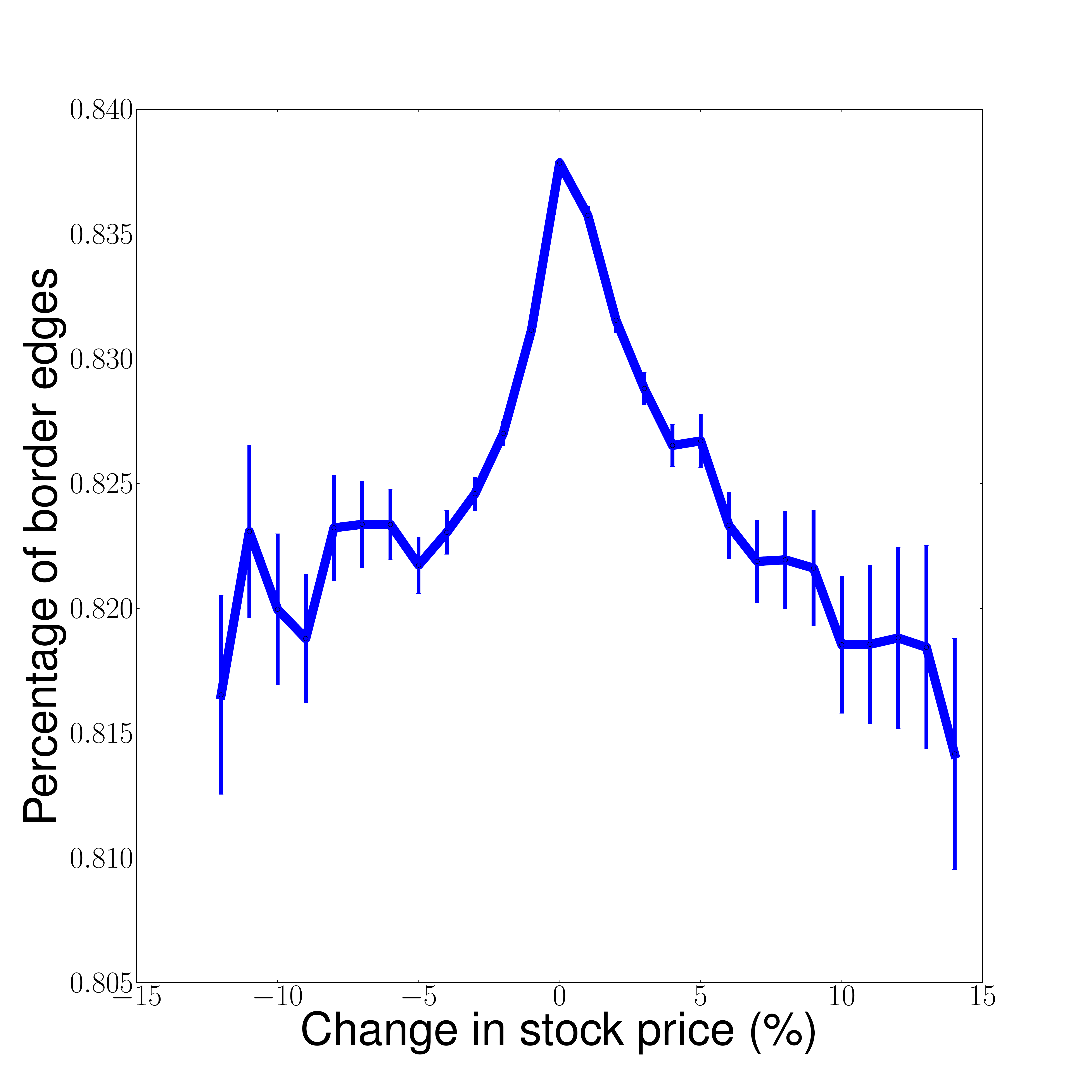} 
\label{RatioBorderToInternalEdges_0_aggregate}
}
\caption{Change in stock prices vs. network structure for networks with two or more nodes.}
\label{N_vs_PriceChange_min2Nodes}
\end{figure*}

We observe a substantial relationship between changes in stock price and changes in network structure.  Figure \ref{N_vs_PriceChange_min2Nodes} shows how each network feature --  clustering, tie strength, and the relative intensity of insider to outsider contacts -- changes with $\Delta_{s,d}$. The horizontal axis of each figure shows the percentage price change $\Delta$. When $\Delta > 0$ ($\Delta < 0$): the vertical axis indicates the mean measure for networks $G_{s,d}$ such that $\Delta_{s,d} \geq \Delta$ ($\Delta_{s,d} \leq \Delta$). When $\Delta = 0$, the vertical axis indicates the mean measure for all networks. Throughout the paper, figures also include 95\% confidence intervals. Throughout the paper, figures also include 95\% confidence intervals. Aggregating changes in network properties with respect to price changes over all stocks and days, we observe that as price changes increase in intensity, the network exhibits a higher clustering coefficient\footnote{We observe consistent results when we measure clustering while controlling for nodes ($\nu(C_{s,d})$) and edges ($\epsilon(C_{s,d})$).}.  Price changes are also related to increasing levels of tie strength\footnote{In Figure \ref{N_vs_PriceChange_min2Nodes}\subref{Strength_of_ties_2_FracTopConnection_aggregate_ComputeEachUser}, we use $\alpha = 0.1$ as the threshold for tie strength.  Border edges also drop as a percentage of edges in the face of price changes.  Repeating the test with various values of $\alpha$ showed that the results are similar regardless of the choice of $\alpha$}. These findings reveal that decision-makers in the network tend to \emph{turtle up} their communication in the face of stress, rather than \emph{open} up.

Changes in connectivity are also significant but substantively slight. The fraction of nodes in the largest component increases slightly and the number of components needed to account for 90\% of the nodes decreases slightly with price changes. We also find a trend with respect to network size --- as the price changes increase, the number of nodes ($\bar{N}_{s,d}$) and edges increases ($\nu(E_{s,d})$).  These effects are reported for graphs $G_{s,d}$ with at least two nodes; all results are consistent if we restrict to graphs containing a larger number of nodes.  We note that changes in network structure are symmetric with respect to the direction of the price change, which is curious in nature given that one might expect that price increases and decreases would be associated with different structural changes --- an issue we address below when we examine the actual changes in actors' cognition and affect.

We further examined the results in Figure \ref{N_vs_PriceChange_min2Nodes} by disaggregating the analysis on a stock-by-stock and industry-by-industry basis subject to control variables. We run OLS regressions of the form $f(G_{s,d}) = \beta_0|\Delta_{s,d}| + \beta_1f(G_{s,d-1}) +  \beta_2f(G_{s,d-2}) + \beta_3VIX + \alpha_1D_s^1 + \dots + \alpha_SD_s^S + r_{s,d}$, where $S$ is the number of stocks in our data set. For each stock $s'$ the indicator function $D_s^{s'}$ is defined as follows as follows:\\

 \begin{displaymath}
   D_s^{s'} := \left\{
     \begin{array}{lr}
       1 & : s = s'\\
       0 & : s \neq s'
     \end{array}
   \right.
\end{displaymath} 

Including this indicator function as an independent variable measures the association between the change in stock prices, $|\Delta_{s,d}|$ and the network properties $f(G_{s,d})$ on a stock-by-stock basis (i.e., fixed effects model \cite{Bhargava}).  We included fixed effect variables for day of the week to control communication patterns explained by the day of the week (e.g., earning announcements and quarterly reports typically are made on specific days for specific stocks); a variable measuring the daily market wide volatility using the VIX index; and the lagged values $f(G_{s,d-1})$ and $f(G_{s,d-2})$ to control for possible autocorrelation in the time series of network properties.

We run regressions with the described stock level fixed effects and an additional version where we use a fixed effects model at the industry level instead of the stock level. Tables \ref{regressions_fe_stock} and \ref{regressions_fe_industry} show the value and significance of each independent variable using fixed effects at the stock and industry level, respectively. The results confirm the analyses presented in Figure \ref{N_vs_PriceChange_min2Nodes}.  Changes in price are associated with increases in network size, clustering, and strength of ties, and decreases in the percentage of outside contacts.  A post-test analysis for autocorrelation of the residuals was also run.  For each stock $s$, we ran a separate regression of the form $f(G_{s,d}) = \beta_0|\Delta_{s,d}| + \beta_1f(G_{s,d-1}) +  \beta_2f(G_{s,d-2}) + r_{s,d}$. Based on the Durbin-Watson test \cite{Durbin} there was no statistical evidence of positive or negative serial correlation in 99.2\% and 99.9\% of the stocks, respectively.

\begin{figure*}
\centering

\subfigure[Average clustering coefficient normalized by number of nodes ($\nu(C_{s,d})$)]{
\includegraphics[width =.31 \textwidth]{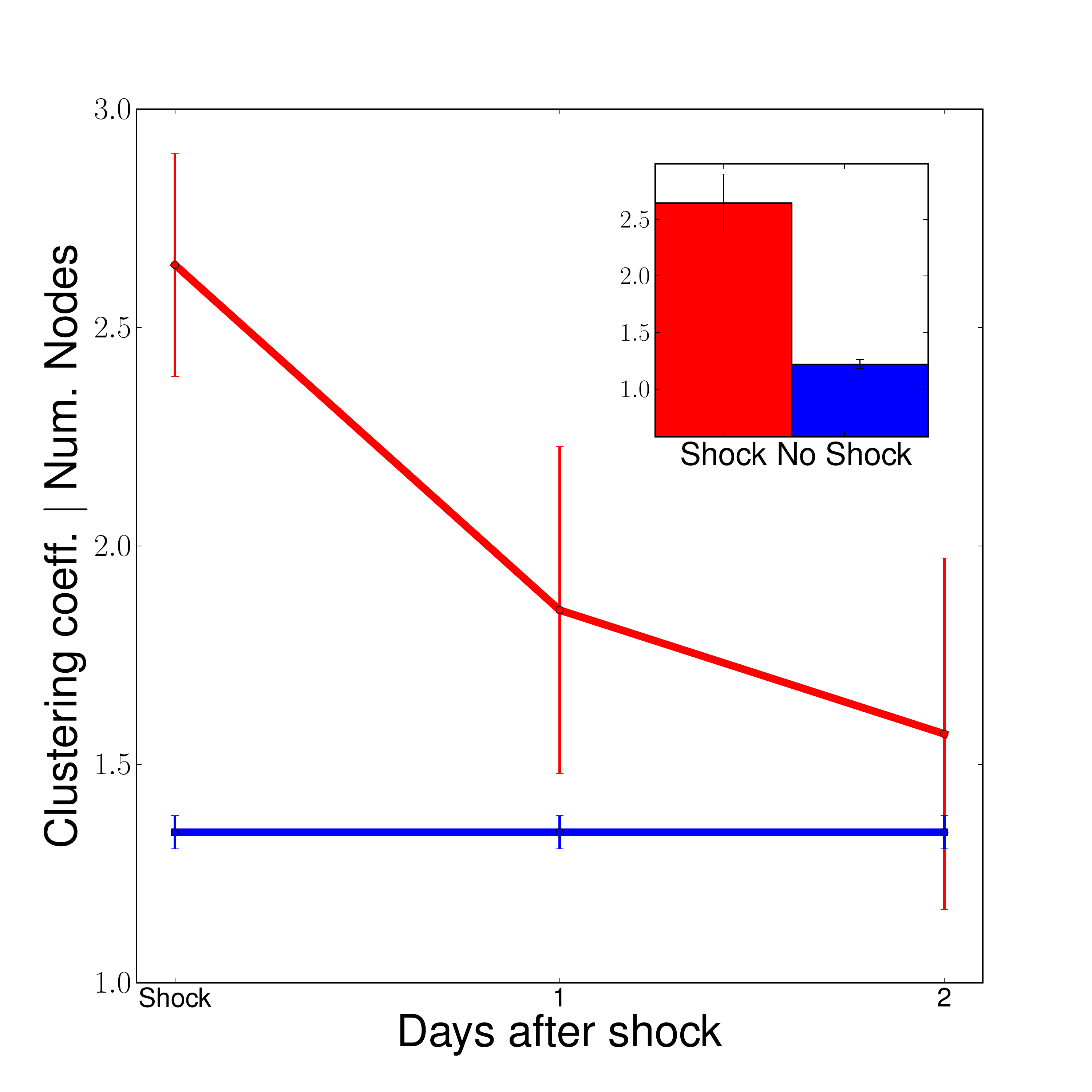} 
\label{clusteringCoeff_cNodes_AllEdges_AfterShock_singletonsAllowed}
}
\subfigure[Strength of ties ($S_{d,s,.1}$)]{
\includegraphics[width =.31 \textwidth]{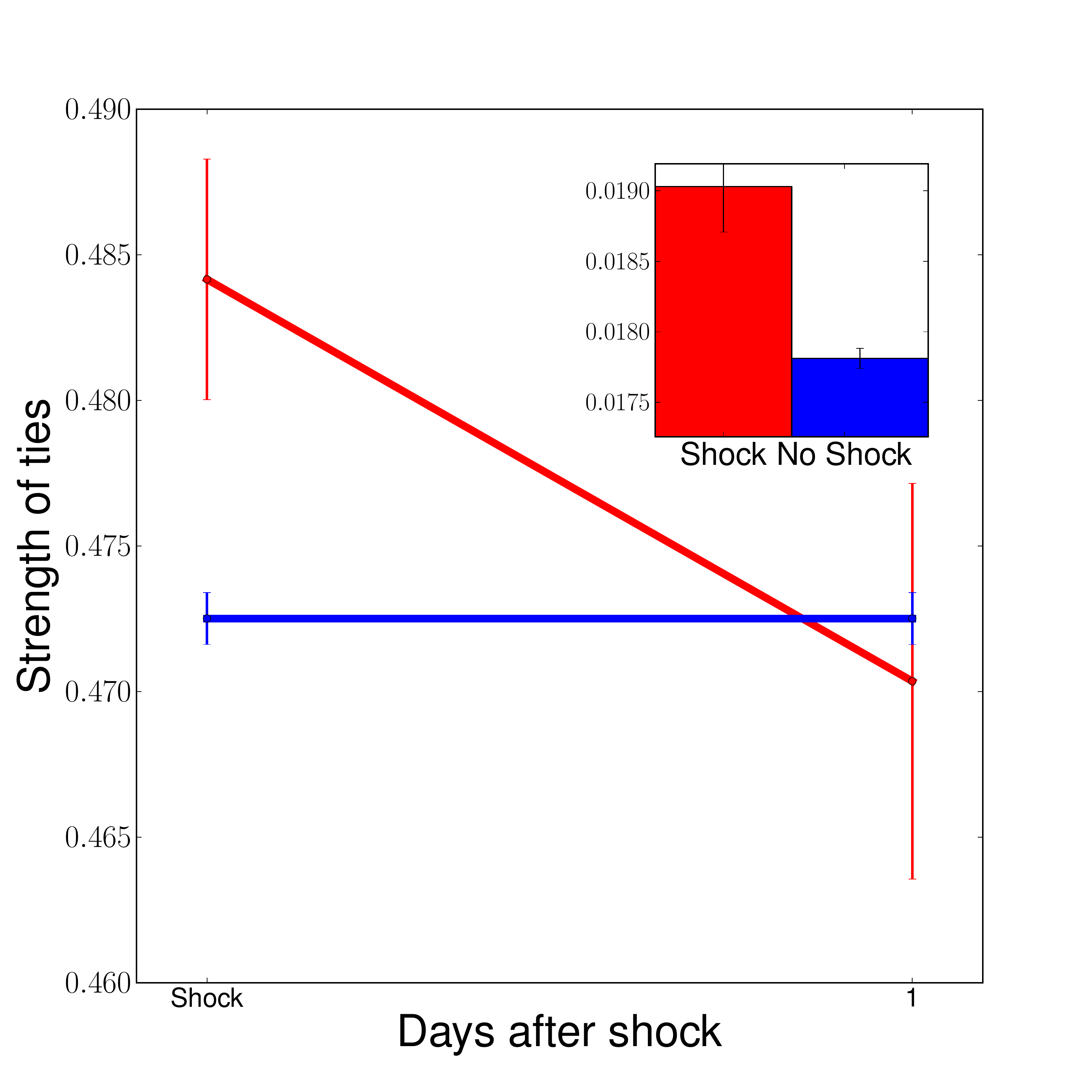} 
\label{Strength_Of_ties_AfterShock_singletonsAllowed}
}
\subfigure[Percentage of border edges ($O_{d,s}$)]{
\includegraphics[width =.31 \textwidth]{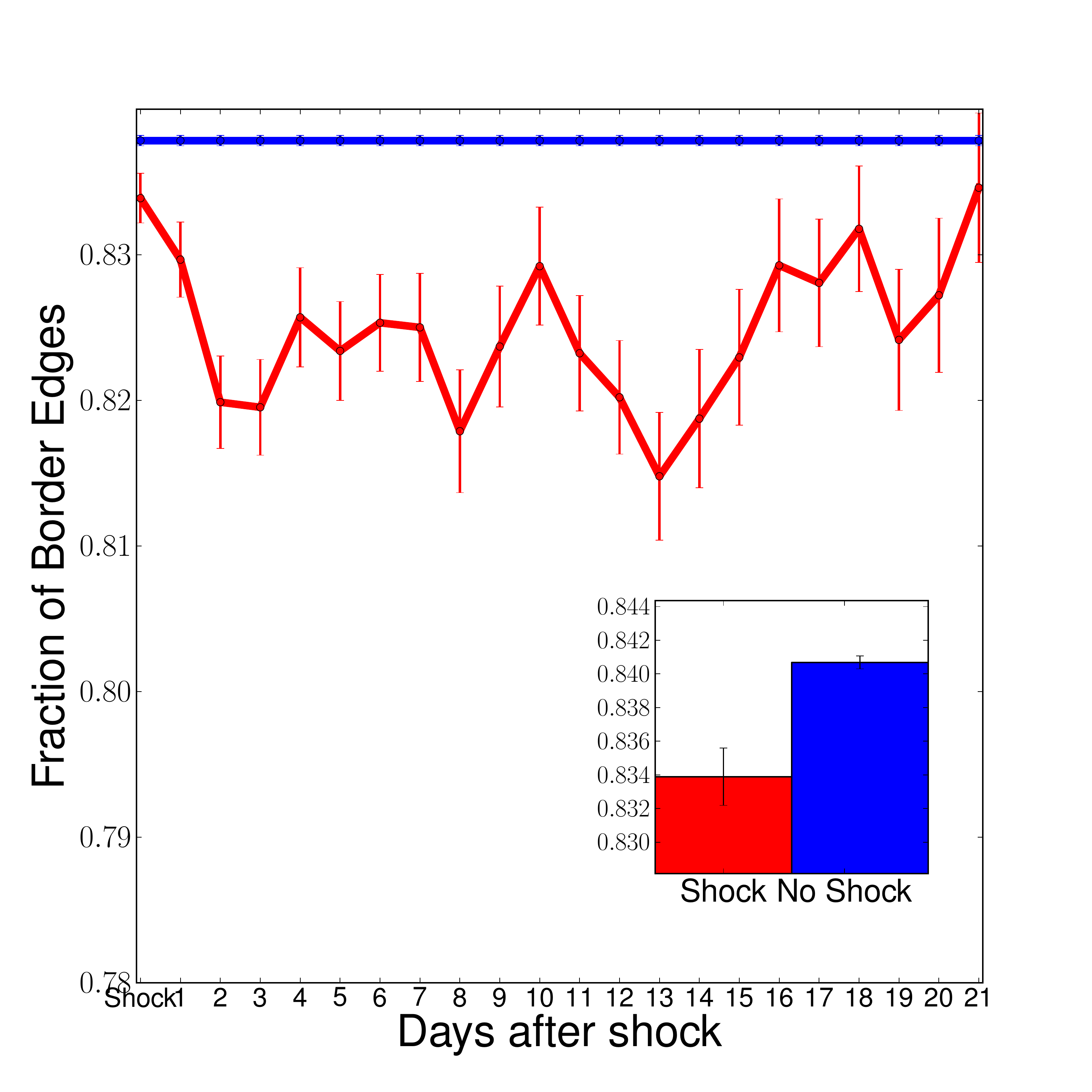} 
\label{RatioBorderToInternalEdges_AllEdges_AfterShock_singletonsAllowed}
}
\caption{Network structure on days with a $x$-shock and on the days following a an $x$-shock.  Insets compare network properties on shock and non-shock days. $x = 5\%$ was used to define $x$-shocks.}
\label{networkFeatures_Shocks}
\end{figure*}

The relationship between unexpected price changes and network structure further substantiate the inference that social networks in the face of uncertainty, as measured here, exhibit a propensity to turtle up.  We compare the value of each feature of $G_{s,d}$ when $(s,d)$ is an $x$-shock and when it is not.  Further, we measure the number of days a graph feature takes to return to its mean value following a shock.  Figure \ref{networkFeatures_Shocks} shows the values of network features in graphs $G_{s,d}$ on the day of an $x$-shock ($x=5\%$, as defined above), and then on subsequent days until the feature approximately returns to its average over all networks.  The insets in Figure \ref{networkFeatures_Shocks} compare network values on shock and non-shock days.  We find that all features are significantly different when there are shocks, and the direction of the differences mirror the previous Figure \ref{N_vs_PriceChange_min2Nodes}.  The same patterns arise when different values of $x$ are used.   Our analysis of network recovery indicates that most network properties return to their average value one or two days after the shock, suggesting that normalization in relation to these shocks is relatively fast acting.  

These results suggest that social networks in the face of stress broadly exhibit features associated with turtling up rather than opening up.  Social networks become more intensely interconnected among third parties, rely more on information from strong rather than weak ties, and disproportionately attend to organizational insiders.

\subsection{Cognitive and Affective Content}

To further understand the behavioral ramifications of network structure in the face of shocks, we explored the links between changes in network structure and the psychology of the actors in the network. If changes in network structure are associated with actual changes in behavior, changes in network structure should predict changes in actors' cognition and affect.  Cognition and affect are important psychological conditions, shaping information processing and decision-making \cite{Kahneman, Tetlock, Agarwal}.  

To infer affect and cognition of the actors in our network, we use the Linguistic Inquiry and Word Count (LIWC) dictionary, which identifies words in the content of communications (i.e., instant messages) that reflect affective and cognitive states\footnote{See http://www.liwc.net for more information on LIWC 2007}. Affect includes positive emotion, negative emotion, anxiety, anger, and sadness.  Cognition includes insight, causation, discrepancy, tentative, certainty, inhibition, inclusive, and exclusive.  the LIWC dictionary is well-validated and regularly used \cite{Kramer, Coviello}. 

\begin{figure*}
\centering
\subfigure[Cognitive processes]{
\includegraphics[width =.31 \textwidth]{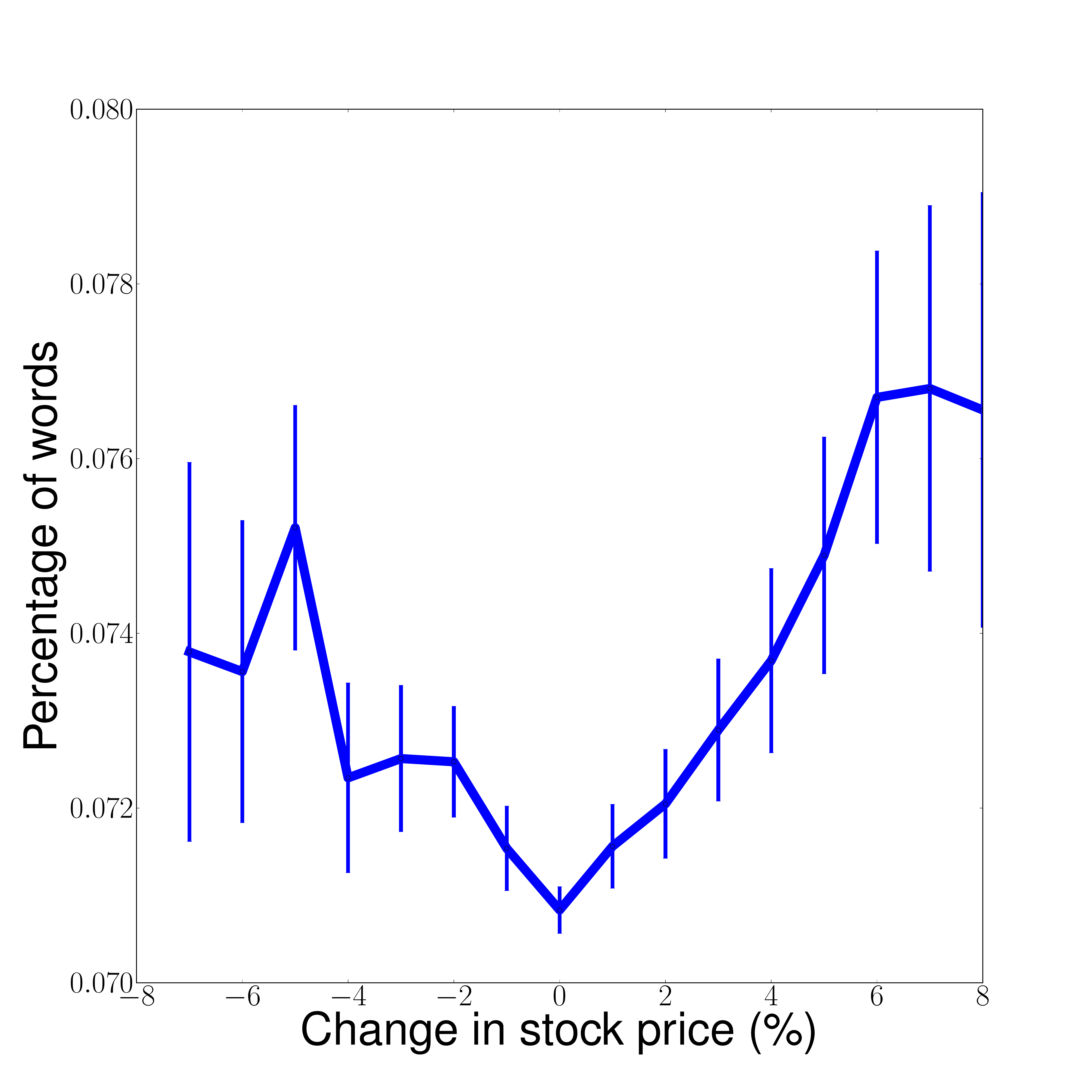} 
\label{Cognitive_processes}
}
\subfigure[Insight]{
\includegraphics[width =.31 \textwidth]{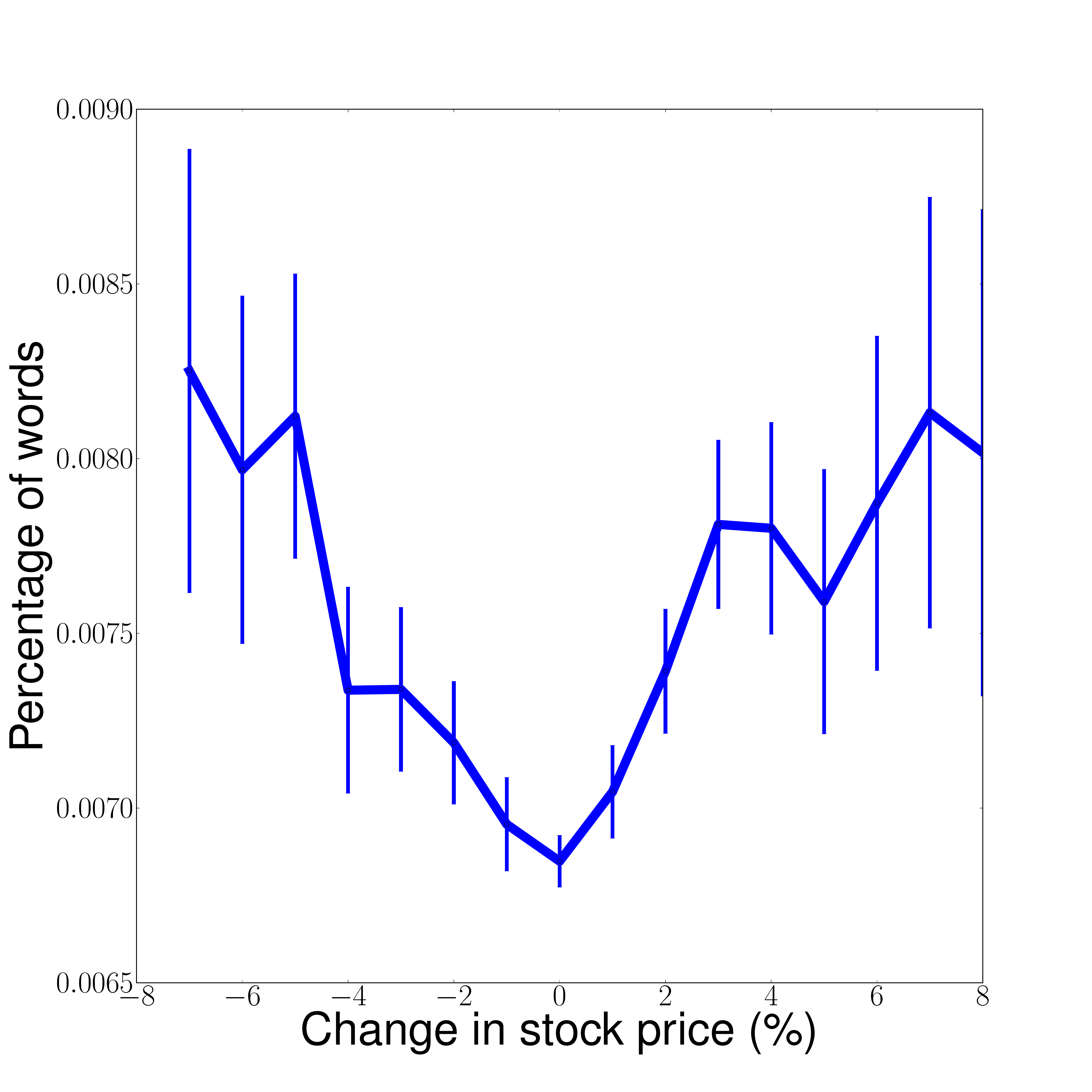} 
\label{Insight}
}
\subfigure[Causation]{
\includegraphics[width =.31 \textwidth]{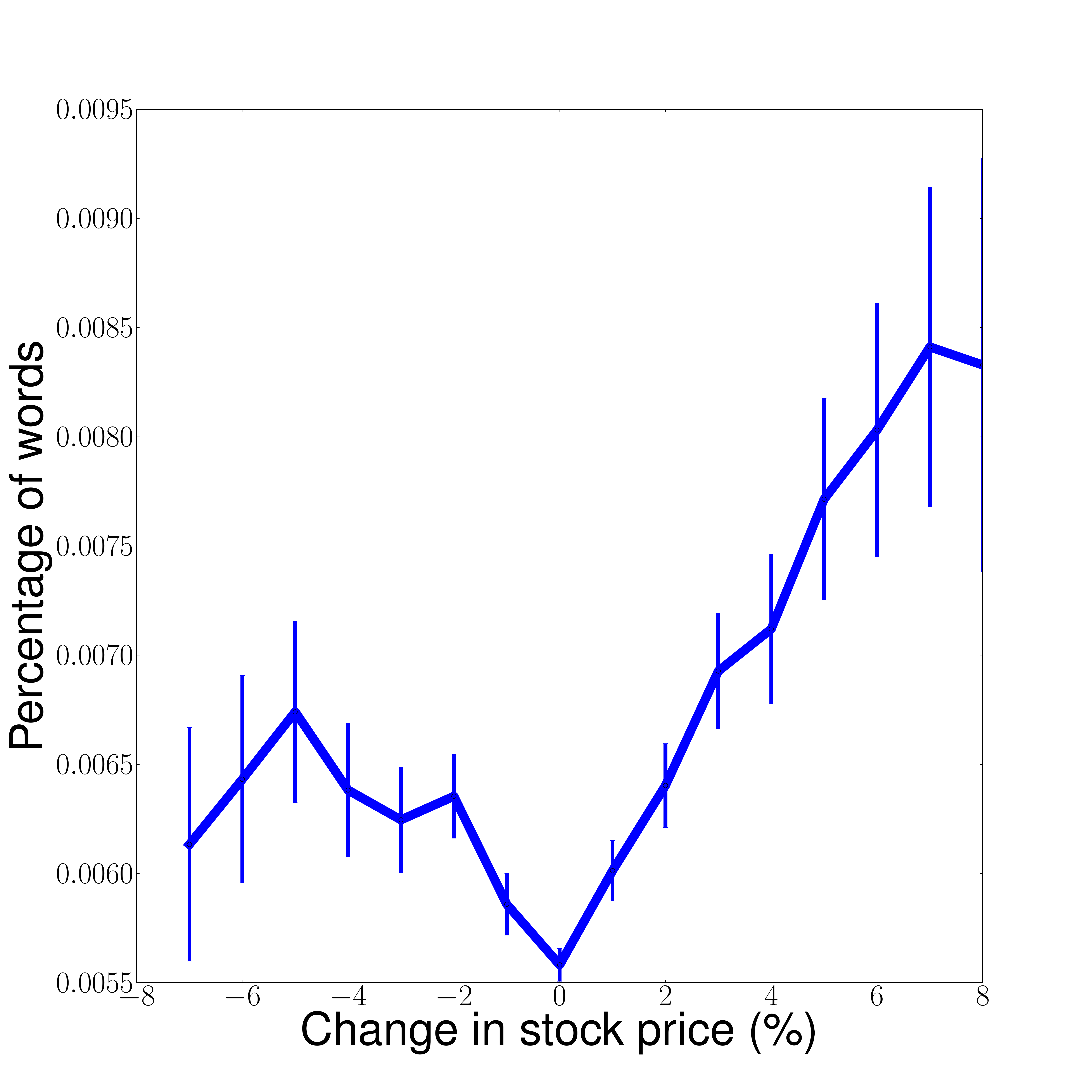} 
\label{Insight}
}
\subfigure[Certainty]{
\includegraphics[width =.31 \textwidth]{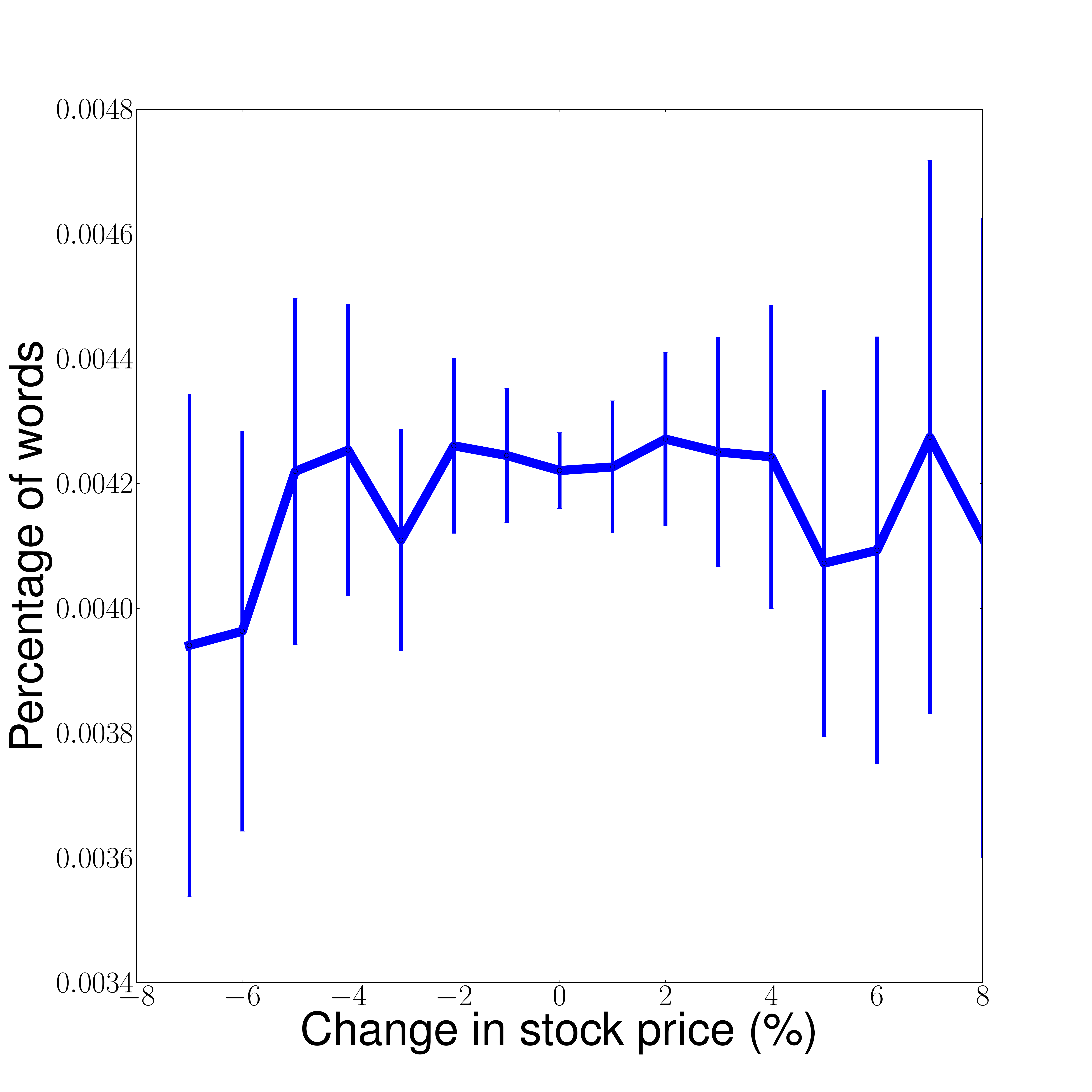} 
\label{Insight}
}
\subfigure[Discrepancy]{
\includegraphics[width =.31 \textwidth]{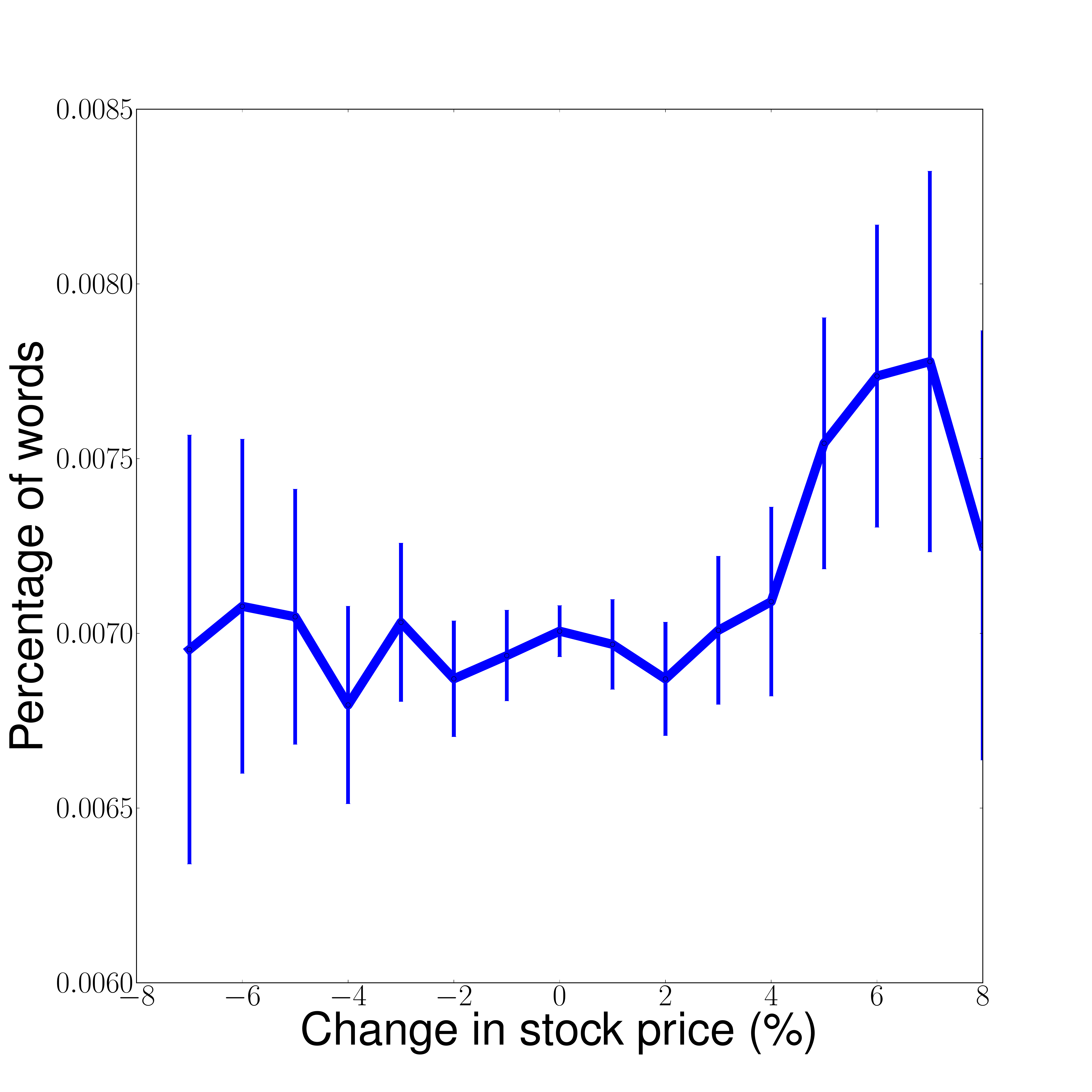} 
\label{Insight}
}
\subfigure[Exclusive]{
\includegraphics[width =.31 \textwidth]{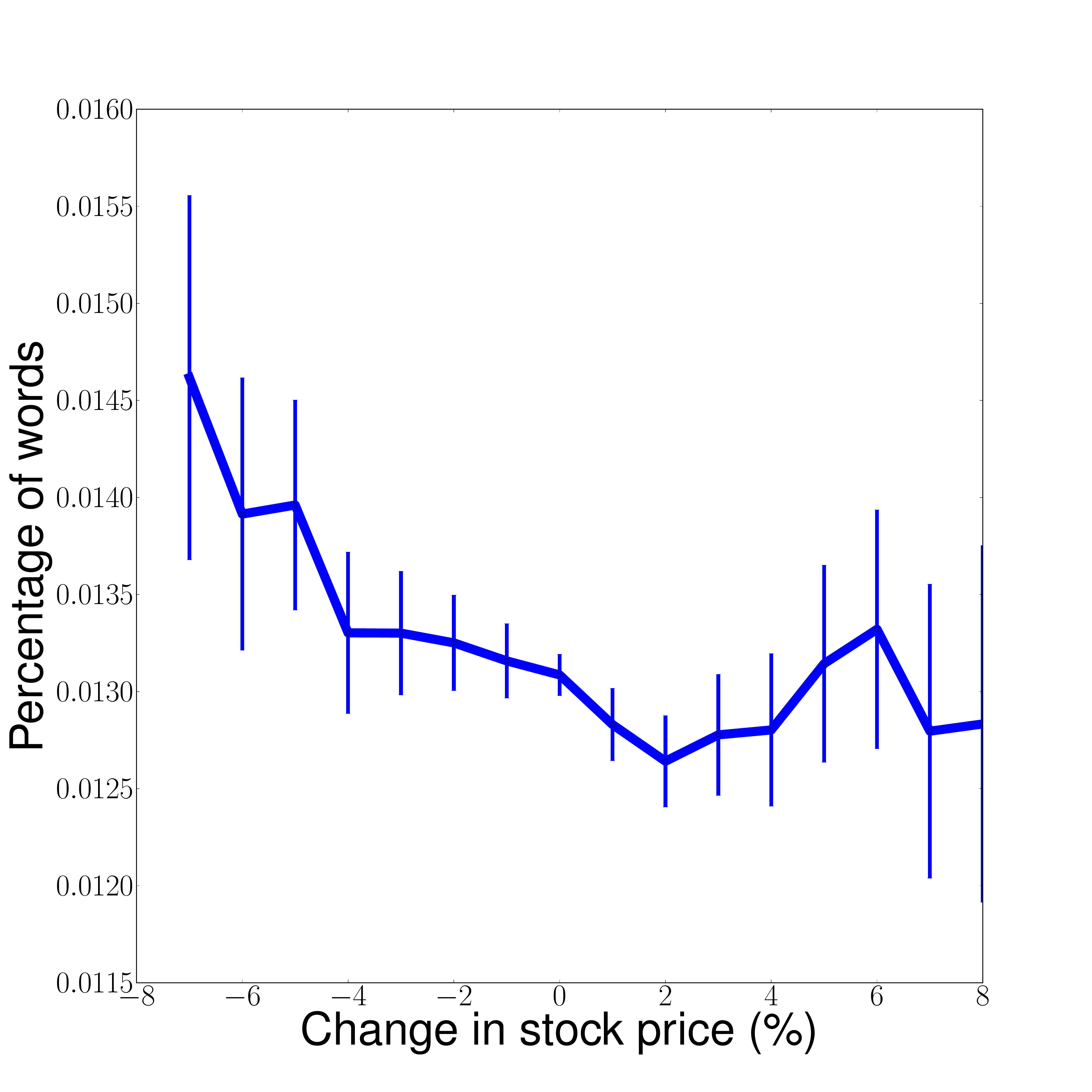} 
\label{Cognitive_processes}
}
\subfigure[Inclusive]{
\includegraphics[width =.31 \textwidth]{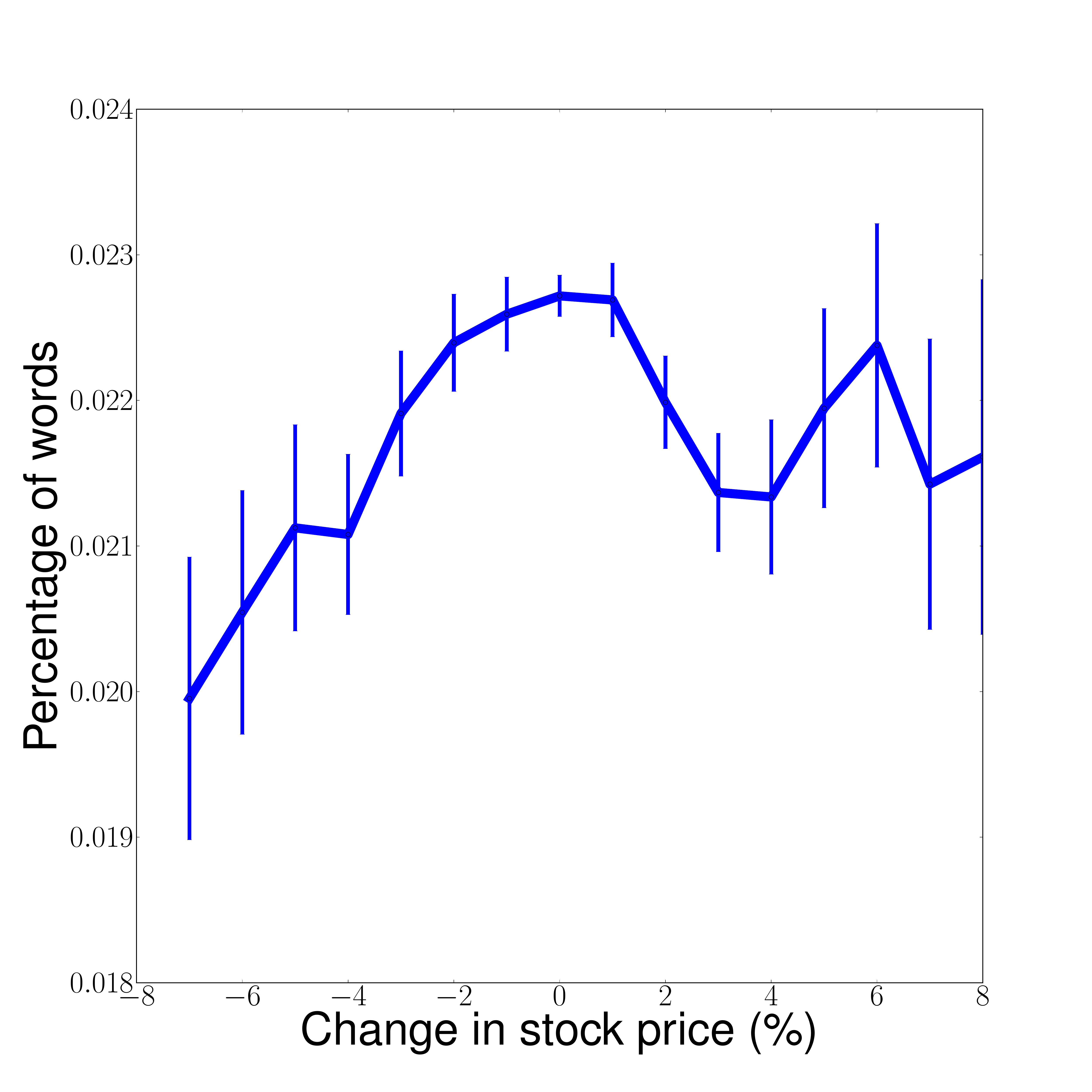} 
\label{Insight}
}
\subfigure[Inhibition]{
\includegraphics[width =.31 \textwidth]{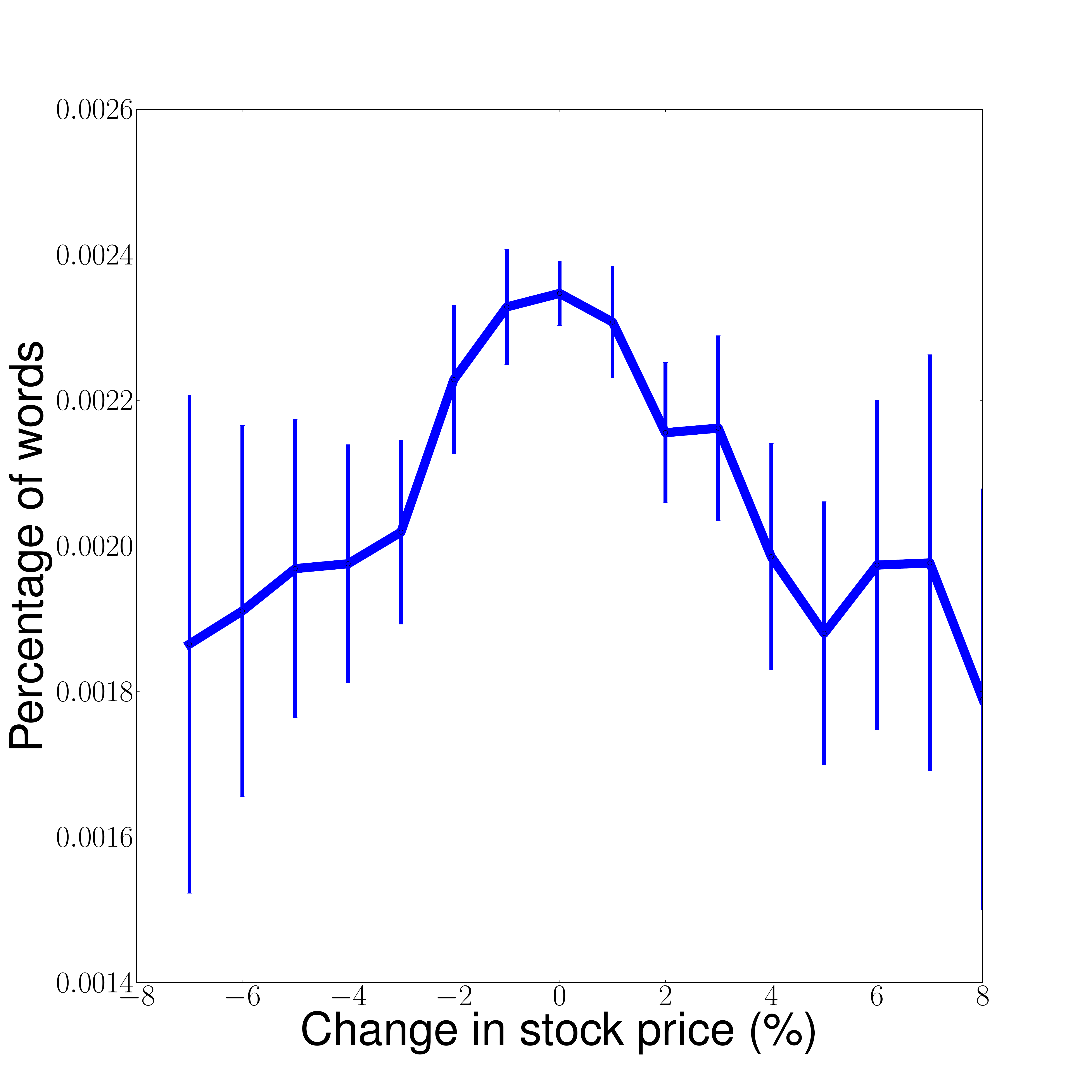} 
\label{Insight}
}
\subfigure[Tentative]{
\includegraphics[width =.31 \textwidth]{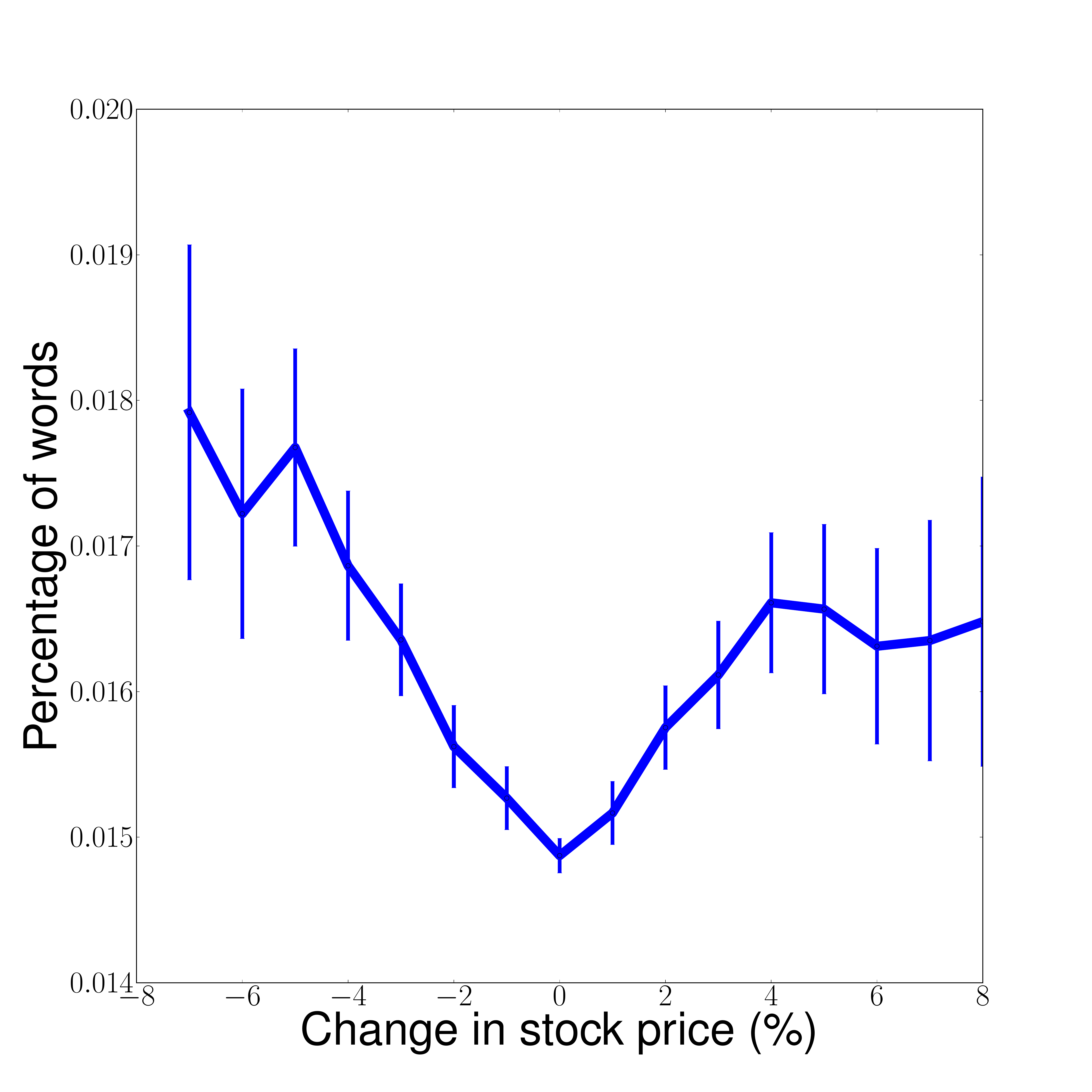} 
\label{Insight}
}

\caption{Price changes vs. percentage change in words reflecting various cognitive processes. \label{CognitiveCat_vs_priceChange}}
\end{figure*}

\begin{figure*}
\centering
\subfigure[Affective processes]{
\includegraphics[width =.25 \textwidth]{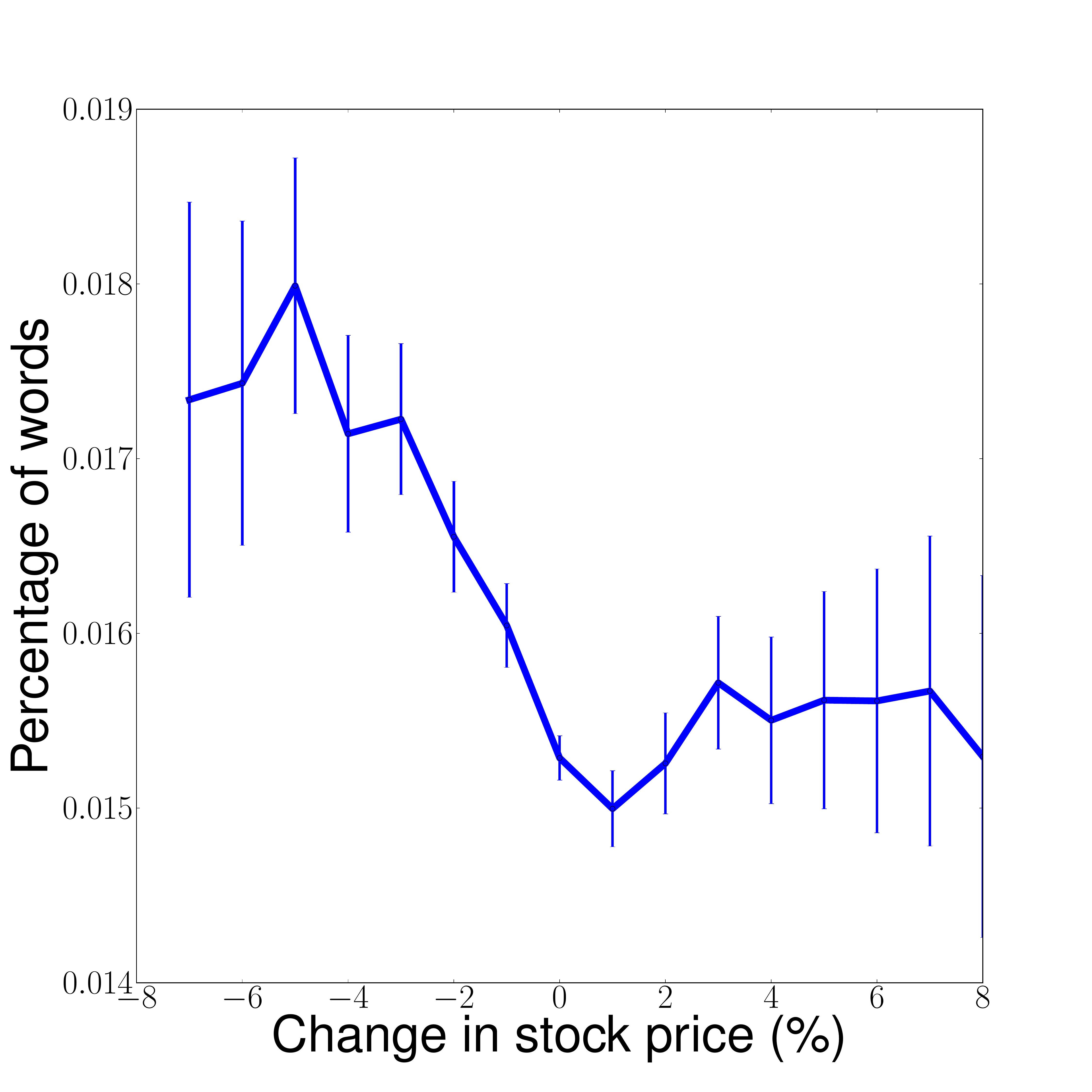} 
\label{Affective_processes}
}
\subfigure[Negative emotion]{
\includegraphics[width =.25 \textwidth]{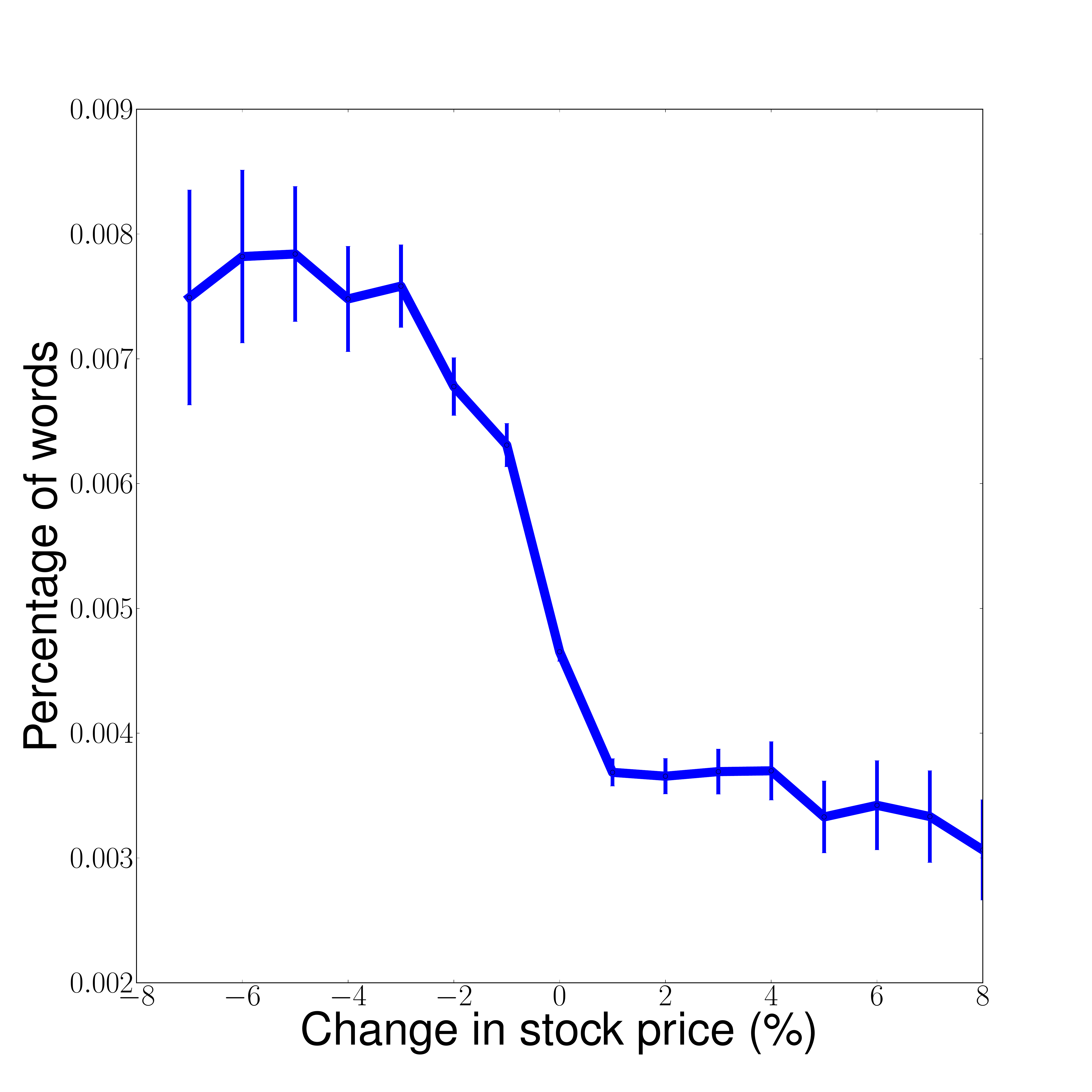} 
\label{Negative_emotion}
}
\subfigure[Positive emotion]{
\includegraphics[width =.25 \textwidth]{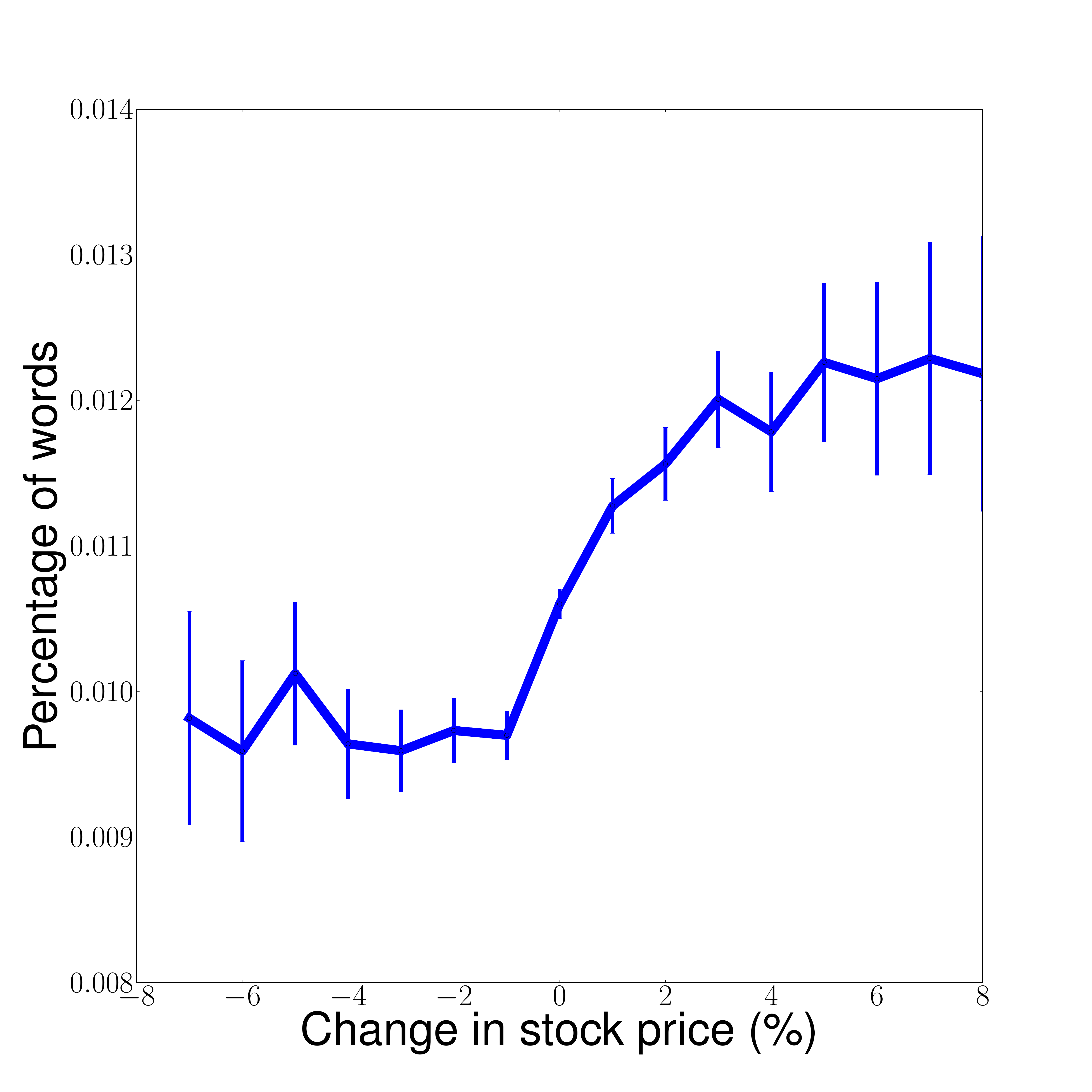} 
\label{Positive_emotion}
}

\subfigure[Anger]{
\includegraphics[width =.25 \textwidth]{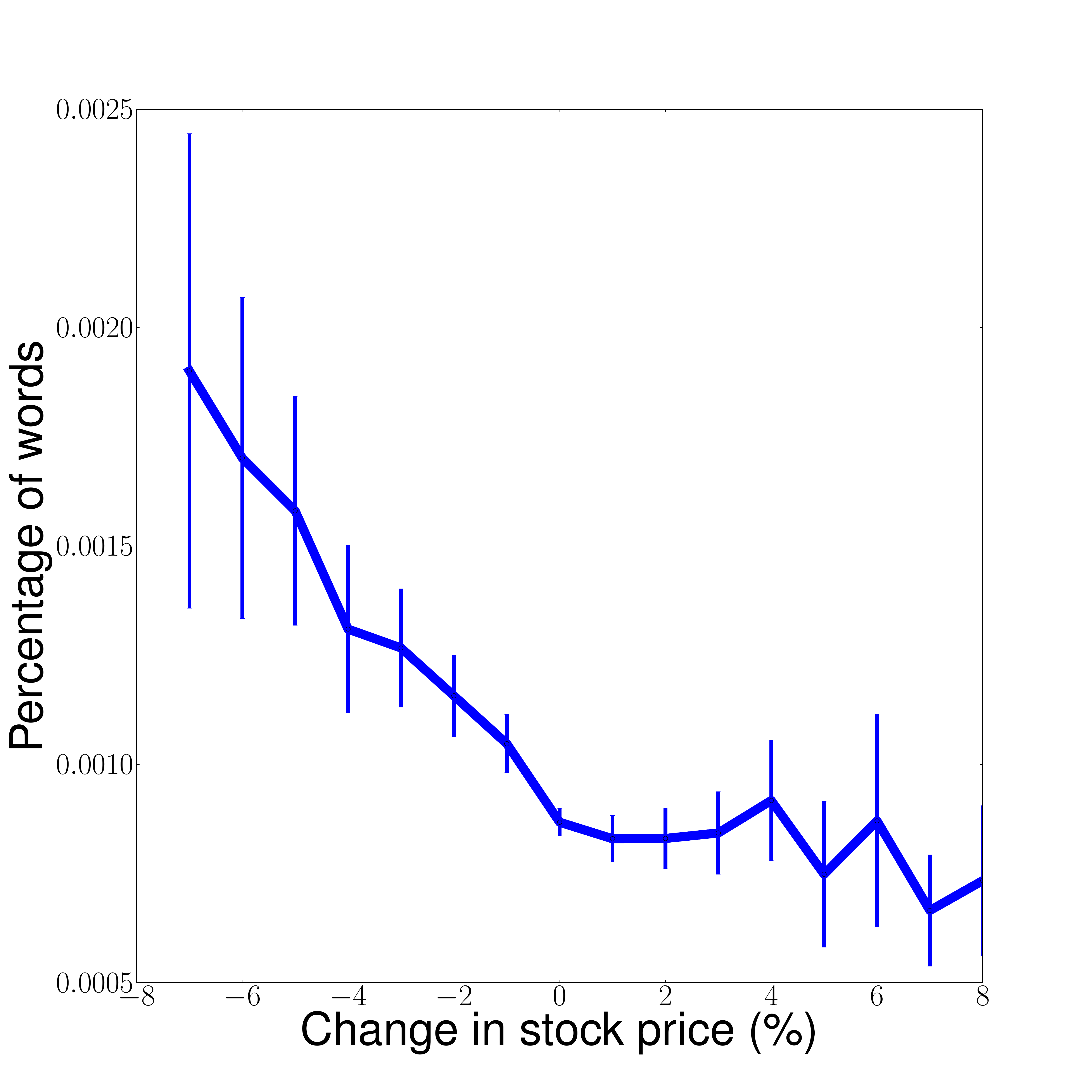} 
\label{Anger}
}
\subfigure[Anxiety]{
\includegraphics[width =.25 \textwidth]{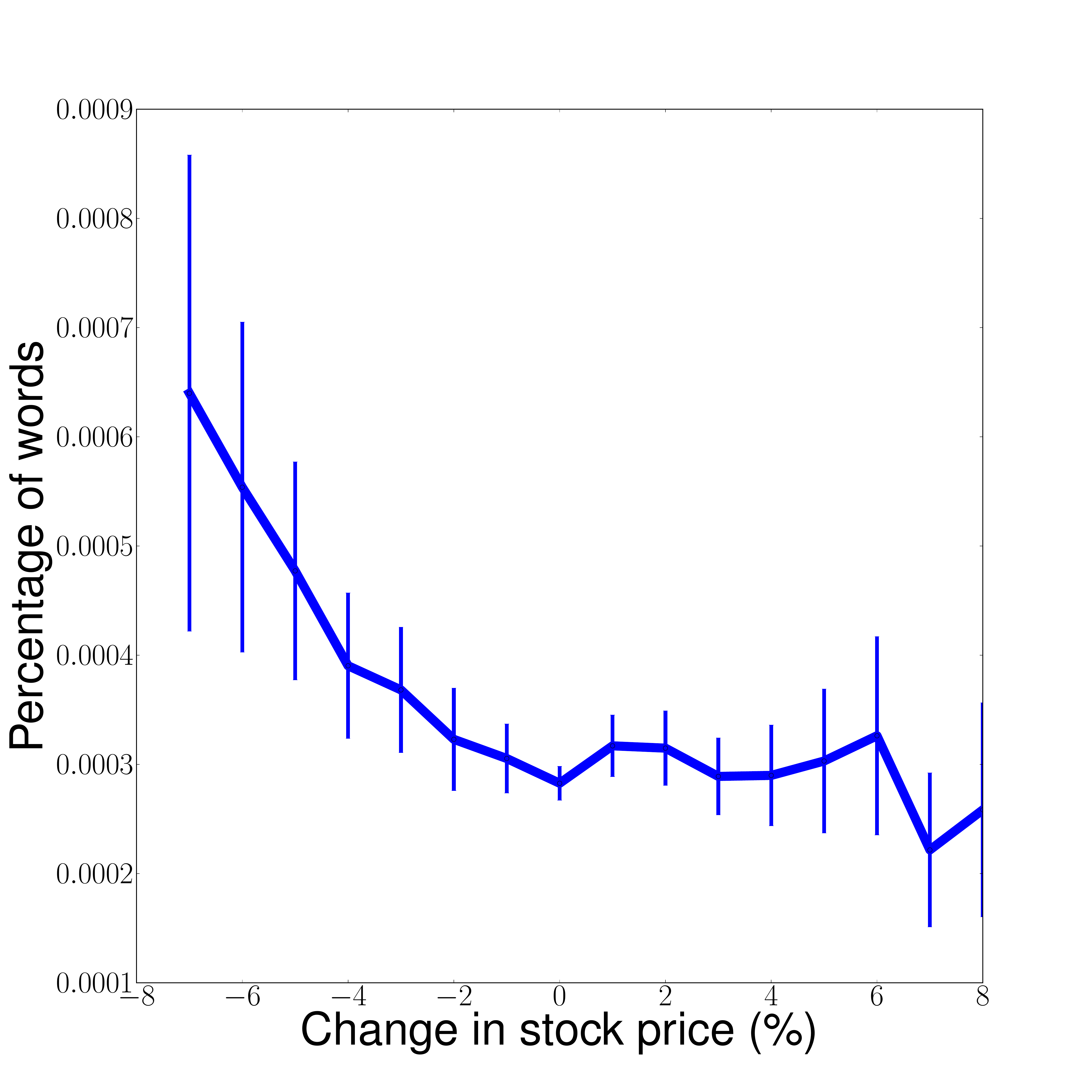} 
\label{Anxiety}
}
\subfigure[Sadness]{
\includegraphics[width =.25 \textwidth]{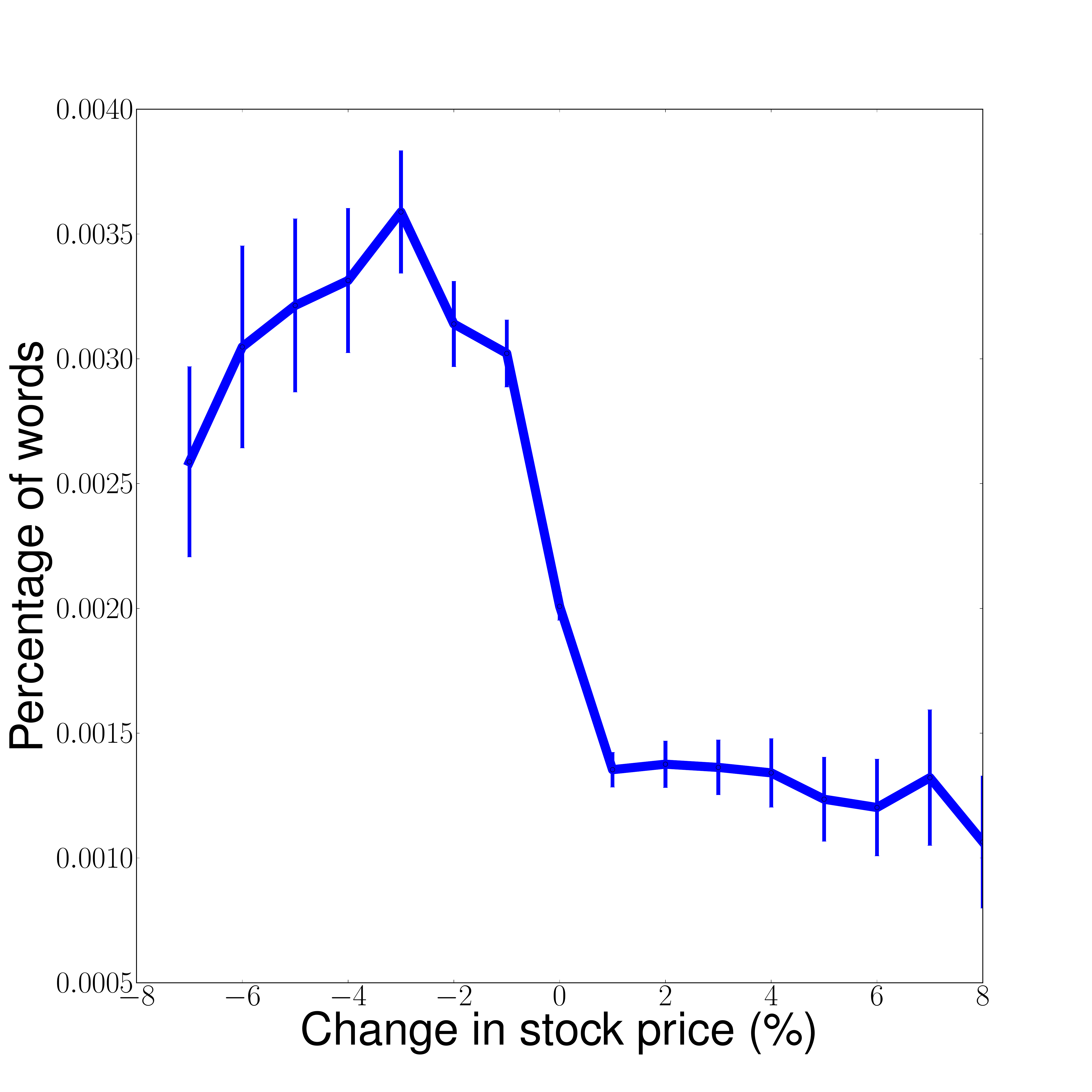} 
\label{Sadness}
}

\caption{Price changes vs. percentage change in words reflecting various affective processes \label{AffectiveCat_vs_priceChange}}
\end{figure*}

Before incorporating the role of network features, we investigate the relationship between stock prices and IM content by measuring how the usage of LIWC categories varies with changes in the stock prices.  In particular, research holds that price changes are stressful for the employees of the firm from a cognitive and emotional perspective \cite{Lo}.  For each instant message among insiders, we computed the percentage of words in each LIWC category.  Formally, given a stock-day pair $(s,d)$ and the IMs that mention $s$ on day $d$, we test whether the percentage of LIWC words in the IMs varies with $\Delta_{s,d}$. The x-axis in Figures \ref{CognitiveCat_vs_priceChange} and \ref{AffectiveCat_vs_priceChange} show the relative price change $\Delta$. When $\Delta > 0$ ($\Delta < 0$) and the y- axis indicates the mean percentage of words in each LIWC category in IMs mentioning $s$ on day $d$ such that $\Delta_{s,d} \geq \Delta$ ($\Delta_{s,d} \leq \Delta$).  When $\Delta = 0$, the y- axis indicates the mean value for all IMs.

We find that price changes are associated with expected changes in cognition and affect.  Figures \ref{Cognitive_processes} shows that as price changes intensify, either upward or downward, decision-makers' communications express higher levels of cognitive processes --- presumably because they face greater risk and complexity in their judgments.  Figures \ref{CognitiveCat_vs_priceChange}(b-i) show how the different subcategories of cognitive processes change with price. The general trend is that the change in cognitive processes is symmetric in the direction of the price change. This symmetric pattern is consistent with the structural changes in Figure \ref{N_vs_PriceChange_min2Nodes}.  

By contrast, affect expressed during ups and downs in price are asymmetric. Figure \ref{Affective_processes} shows that expressed affect surrounding a stock during its price changes differ depending on whether prices rise or fall.  Figures \ref{AffectiveCat_vs_priceChange}(b-f) show the change in the affective subcategories with changes in prices.  Words related to positive emotions are used more when stock prices rise while words related to negative emotions, anger, anxiety, and sadness are used more often when stock prices drop.  One explanation of this is that while funds can make as much money when stock prices fall as when they rise by selling short, more negative affect is expected when prices drop because falling prices generally sound alarms among retail investors who put their capital in the hands of hedge fund decision-makers and take money out of the market when stock prices fall.   

If message content varies with price changes, does network structure further predict the psychology of traders?  

To address this question, we formulate the following prediction task. For each stock $s$ and each LIWC category $C$, let $C_s$ be the fraction of all IMs containing stock symbol $s$ that include a word from category $C$.  For each day $d$, let $C_{s,d}$ be the fraction of all IMs containing stock symbol $s$ on day $d$ that include a word from category $C$. We say that the pair $(s,d)$ {\em conforms} to category $C$ if $C_{s,d} > C_s$ --- in other words, if words from category $C$ are used at a higher rate on day $d$ than is typical for stock $s$.

We use binary classifiers to predict whether each pair $(s,d)$ conforms to each of the cognitive and emotional LIWC categories using the properties of the network and the stock price changes as predictors.  We run the prediction test using three feature sets: only network change features (properties of $G_{s,d}$ described in section 3.1), only price change features ($\Delta_{s,d}$ and $|\Delta_{s,d}|$), and the two sets of features together. Each set of features includes lagged values for 7 days before day $d$.  To test the accuracy of the classifiers, we split time into 100-day bins, using each bin as a test set and all the previous bins as training data.  We balance the testing and training data by including all positive examples and selecting a random sample of negative examples of the same size as the set of control cases. Table \ref{imbalance} shows the number of cases in each class before being balanced.   We use seven binary classifiers: Random Forest, Linear Discriminant Analysis, Quadratic Discriminant Analysis, Naive Bayes, Decision Trees, Support Vector Machines, and Logistic Regression.  Choice of classifier does not change the results; logistic regression results are presented. 

\begin{table}
\centering
\caption{Number of positive and negative examples of pairs $(s,d)$ that conform to each LIWC category.}
\label{imbalance}
\begin{tabular}{c|c|c}
\hline
 \bf{Category} & \bf{Num. Positive} & \bf{Num. Negative} \\
 \hline
 \hline
Affective processes  & 27229 & 65704 \\
\hline
Anger & 3047 & 89886 \\
\hline
Anxiety & 1570 & 91363 \\
\hline
Causation & 14794 & 78139 \\
\hline
 Certainty & 13113 & 79820 \\
 \hline 
 Cognitive processes & 50088 & 42845 \\
 \hline
 Discrepancy & 19281 & 73652 \\
 \hline
 Exclusive & 27383 & 65550 \\
 \hline
 Inclusive & 36033 & 56900 \\
 \hline
 Inhibition & 7556 & 85377 \\
 \hline
 Insight & 19663 & 73270 \\
 \hline
 Negative emotion & 11478 & 81455 \\
 \hline
 Positive emotions & 23551 & 69382 \\
 \hline
 Sadness & 4531 & 88402 \\
 \hline
 Tentative & 27784 & 65149 \\
 \hline
\end{tabular}
\end{table}

\begin{figure}
\centering
\includegraphics[width =.45 \textwidth]{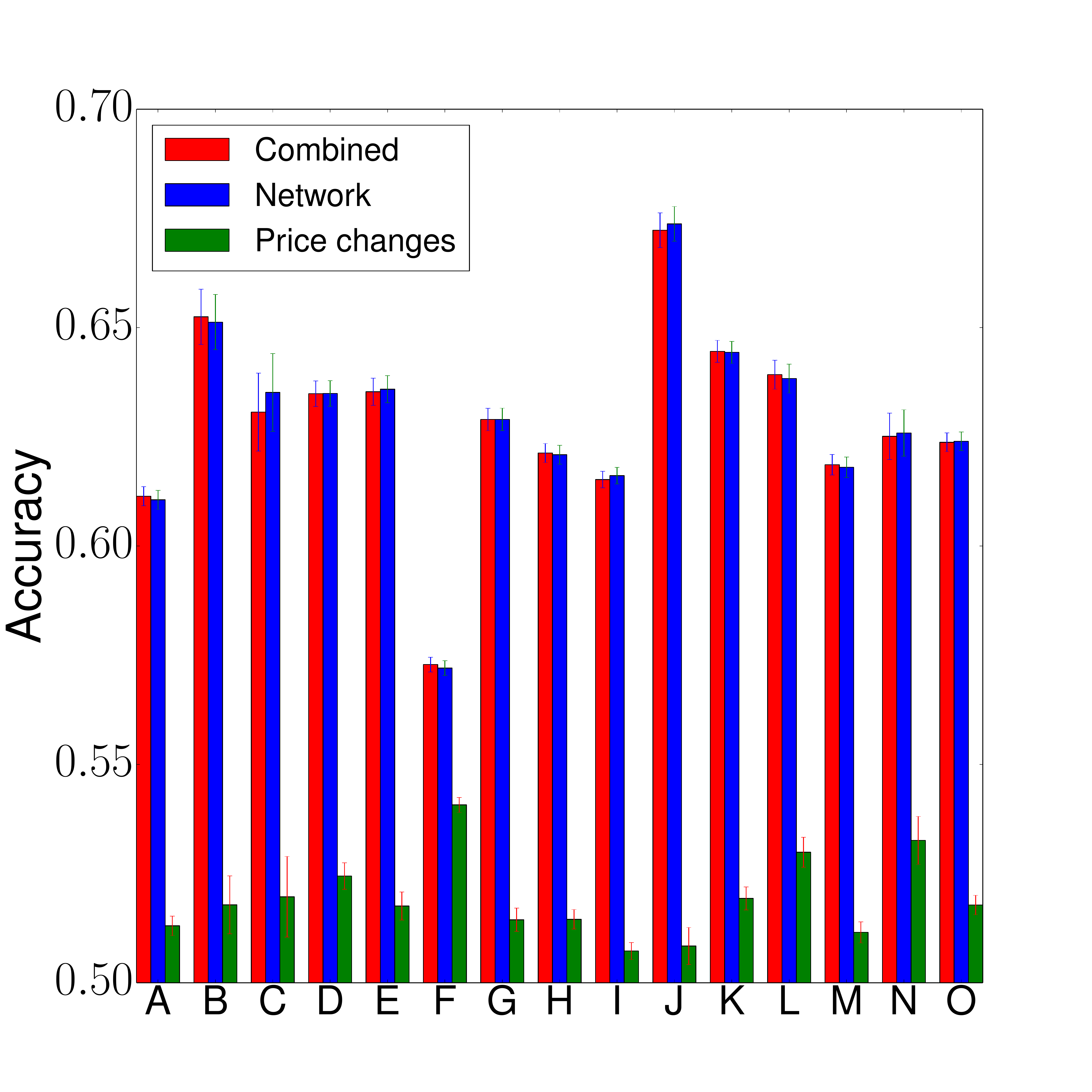} 
\label{accuracy_PredictWordCategory_features_network_priceChanges_catAbveBackgroundRate}
\caption{Prediction accuracy of logistic regression classifier for LIWC categories using network features, price change features, and all features combined. Category key: Affective processes (A), Anger (B), Anxiety (C), Causation (D), Certainty (E), Cognitive processes (F), Discrepancy (G), Exclusive (H), Inclusive (I), Inhibition (J), Insight (K), Negative emotion (L), Positive emotions (M), Sadness (N), Tentative (O).  \label{accuracy_behaviours}}
\end{figure}

Figure \ref{accuracy_behaviours} shows the accuracy of the logistic regression classifier using each of the feature sets for the affective and cognitive categories, and for the positive-emotion, negative-emotion, and insight subcategories.  Network properties alone provide significantly better predictions of the psychology of the decision-makers than the price changes alone. Furthermore, the figure indicates that combining the two types of features does not yield a significant improvement over the network features alone.  This pattern suggests that network structure is more predictive of a network's collective affect and cognition than the price shock itself.

\subsection{Decision Making Behavior}

We now turn to the question of whether the structure of the communication networks provides insight about the trading behavior of the firm. Since trading requires coordination and communication among the employees, it is likely that some of the high level trading decisions of the firm are latent in the structure of the firm's communication. We look at two important aspects of the decision making process when the firm decides to make a trade -- the quality of its trading decision in terms of timing, and the decision to begin trading a stock that has not been traded for a long period of time. Stock prices are crucial in both trading time and in deciding to trade a new stock, hence when we test the predictive value of the networks, we always compare it against the stocks' price changes. We find that the network structure is indeed predictive of these trading behaviors, above and beyond what is predicted by price changes alone.

\subsubsection{Predicting Performance.} 
We begin by measuring whether the timing of each trade
was \emph{locally optimal}. To measure local optimality of each transaction,
we ask whether the firm would have benefited from waiting until the
next day to make the transaction. For a stock $s$ traded on day $d$ we
let $p_{s,d}$ denote the price of the stock for that transaction. We
let $p^{max}_{s,d}$ and $p^{min}_{s,d}$ be the maximum and minimum
price of stock $s$ on day $d$. If a stock $s$ was bought on day $d$ at
price $p_{s,d}$ and the maximum price of $s$ the following day
($p^{max}_{s,d+1}$) was less less than $p_{s,d}$, then the company
would have benefited from waiting until the next day to buy stock $s$.
Following this reasoning, we label each \emph{buy} transaction
$(s,d,p_{s,d})$ as \emph{locally suboptimal} if $p_{s,d} > p^{max}_{s,d+1}$
and \emph{locally optimal} otherwise. 
Similarly, we label each \emph{sell}
transaction $(s,d,p_{s,d})$ as locally {suboptimal} if $p_{s,d} >
p^{min}_{s,d+1}$ and locally {optimal} otherwise. We observe that about
20\% and 23\% of all buy and sell transaction are locally suboptimal.

We expect that the decision process of the firm does not always
involve optimizing exactly the day on which to make a transaction. Instead,
the firm may focus most of its efforts on other objectives --- trying to decide
\emph{which} stocks to trade, \emph{how much} to trade, or on long term
profit as opposed to small gains. To validate our measure of
local optimality, we test whether the company's trades reflect an effort to
trade on a locally optimal day by comparing the firm's real performance
with their performance if the trades occurred on a random day.

We create a random set of transactions by taking each transaction in the data set and generating an alternative transaction of the same stock and same number of shares, but on a randomly selected day. The price of the alternative transaction is selected uniformly at random between the minimum and maximum price on the selected day. 

The number of locally
suboptimal transactions in the random set is 2\% higher than
in the actual set of transactions. While 2\% may not initially 
appear to be very large, it is easier to assess the size of the difference when
we compute the loss in profit that these locally suboptimal
transaction generate.
For each locally suboptimal transaction $(s,d,p_{s,d})$, we let
the number of shares traded be $V_{s,d,p}$. The loss generated by this
transaction is $V_{s,d,p}*|p_{s,d} - p_{s,d+1}|$. The difference in
total loss generated by the set of actual transactions and the set of
random transaction is about \$40 million. This shows that the company
performs much better than they would if they traded the same stock and
the same number of shares, but on a randomly selected day. This
validates our measure; we now test if it is related
to features of the networks.

We use our set of classifiers to predict whether transactions are 
locally optimal using the properties of the network and the stock price
changes as predictors. For each transaction $t_{s,d}$ of stock $s$ on day $d$, we use the features of the graph $G_{s,d}$, the chance in
price $\Delta p_{s,d}$, and the absolute change in price $|\Delta
p_{s,d}|$ to predict whether $t_{s,d}$ is locally optimal. 

As we did before, we split time into 100 bins, and use each bin as a
test set and all the previous bins as training data. We also use a balanced set of positive and negative examples.
Since some stocks are traded very often and others are rarely traded, we further split the prediction task into subtasks by the number of
consecutive days on which transactions occurred. Letting $T_k$ be the
set of transactions that have occurred on at least the previous $k$
days, we run the classifiers on each set of transactions $T_k$ for
$k=0 \dots 6$.  

\begin{figure}
\centering
\includegraphics[width =.5 \textwidth]{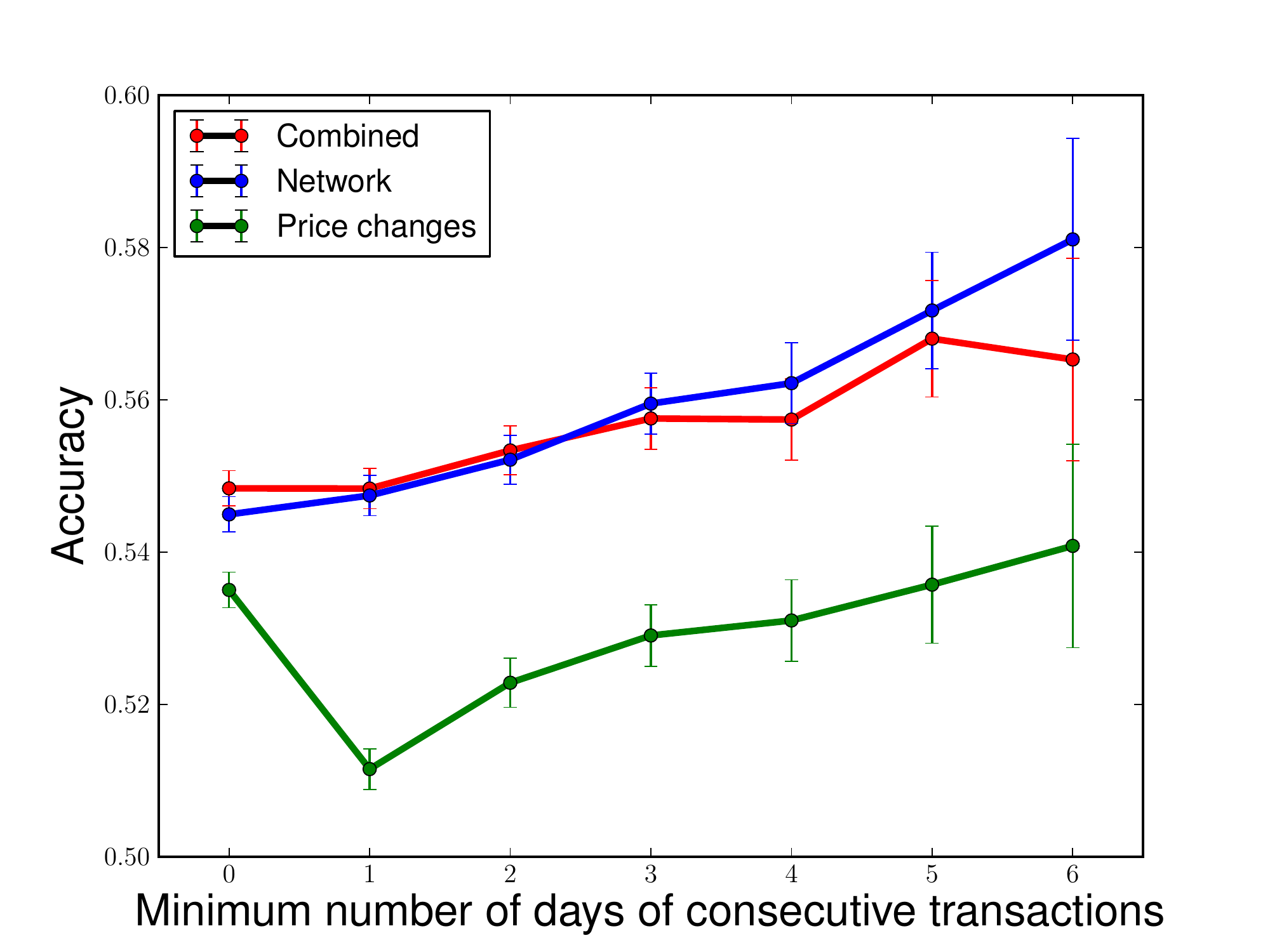} 
\label{NumDaysConsecutiveTrades_vs_accuracy_features_network_pricechange}
\caption{Prediction accuracy of logistic regression classifier when predicting if a trade is locally optimal vs. the number of consecutive days the stock was traded. The curves show the results using network features, price change features, and all features combined. \label{NumDaysConsecutiveTrades_vs_accuracy_features_network_pricechange_optimality}}
\end{figure}

Figure
\ref{NumDaysConsecutiveTrades_vs_accuracy_features_network_pricechange_optimality}
shows the accuracy of the Logistic Regression classifier for each set
of transactions $T_k$. The accuracy of the classifiers increases with
the number of days of consecutive transitions, suggesting
the local optimality of routine transactions is easier to predict than
that of unexpected transactions. We also observe that the network
features are significantly more predictive than the price changes.
Furthermore, combining the network and price change features does not
significantly improve the accuracy of the classifier with network
features alone. This shows that the local optimality of the firm's
decisions are better aligned with the properties of their communication
than the changes in stock prices.

\subsubsection{Predicting Sudden Trading}

We observed that it's easier to predict the local optimality of stocks
that are traded consecutively. We now turn to a second, more basic
question to study the firm-level actions with respect to trading: are
we able to predict whether or not a stock is traded on a given day?
Our first observation is that many stocks are traded at very high
frequencies, and hence the best predictor of a stock being traded on
given day is whether it was traded during the previous few days. 
Given this, we pose the task of predicting 
\emph{new transactions} --  trades of
stocks that have not been traded for a given number of days prior
to the transaction being considered.

We let $NT_k^d$ be the set stocks that have not been traded for $k$ weeks prior to day $d$. We say that a stock $s$ is \emph{k-unobserved} on day $d$ if it is in the set $NT_k^d$. That is, $s$ is $k$-unobserved on day $d$ if it has not been traded during the past $k$ weeks preceding $d$. Note that a $k$-unobserved stock on day $d$ is also $k'$-unobserved on day $d$ for all $k' < k$. We say that all stocks are $0$-unobserved on all days.  

We use our binary classifiers to predict whether stocks that are
$k$-unobserved on day $d$ will be traded on day $d$. The setup of the
prediction task is the same as in the last two sections -- we split
time into 100 bins and use each bin as a test set and all the previous
bins as training data, and we use a balanced set of positive and
negative examples. We use network and price change features as we did
in previous sections, but in this case we also include features that
indicate whether the stock was traded on the 7 days prior to the
$k$-week period when the stock was not traded. For example, when we
predict if a 1-unobserved stock $s$ is traded on day $d$, we know $s$
was not traded on days $d-1, \dots d-7$, but we include features that
indicate if $s$ was traded on days $d-8, d-9, \dots d-14$.

Figure
\ref{NumDaysConsecutiveTrades_vs_accuracy_features_network_pricechange}
shows the accuracy of the classifiers for different values of $k$.
When $k=0$ we are predicting whether stocks are traded on day $d$,
regardless of whether they have been traded on the previous days. We
observe that with no minimum window on the time since the last trade,
the 7-day trading history of the stocks provides over
80\% accuracy, and adding information about the network or stock prices
does not significantly increase the accuracy. However,
as $k$ increases and we begin predicting trading of stocks that have
not been traded for some time, the accuracy of the stocks' trading
history alone drops significantly. When we add the price change
features to the trading history features, we do not observe a
significant increase in accuracy. However, adding the network features
to the trading history features yields a large increase increase in
accuracy. For example, for $k=4, \dots, 9$ the accuracy of the price
change and trading history features is less than 55\% and that of the
network and trading history features is over \%68. Finally, adding 
price change features to the network and trading history does
not significantly increase accuracy. This pattern is consistent
with what we found when predicting affective and cognitive content
and trading performance.

\begin{figure} \centering \includegraphics[width =.5
\textwidth]{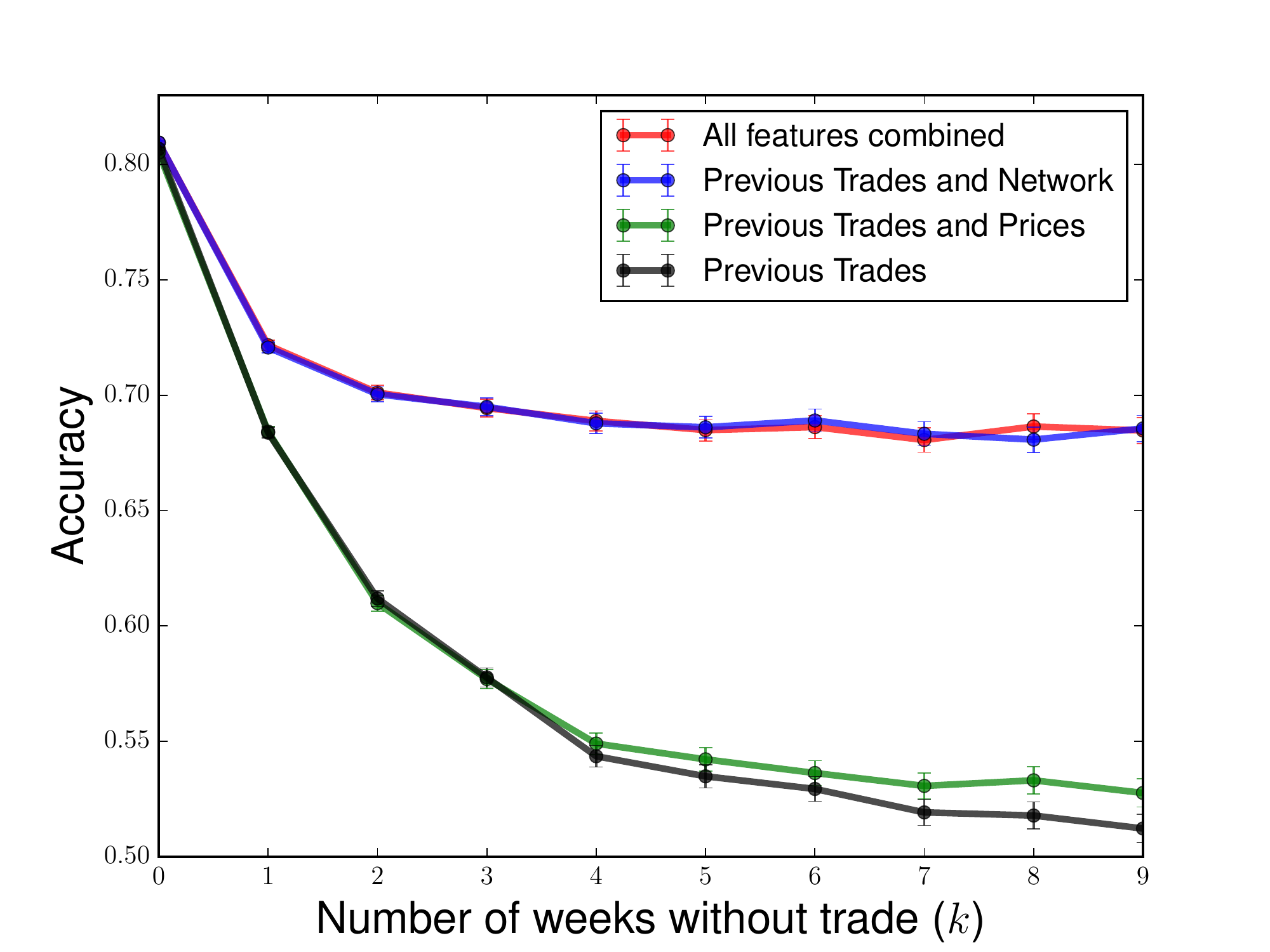}
\label{NumDaysConsecutiveTrades_vs_accuracy_features_network_pricechange}
\caption{Prediction accuracy of logistic regression classifier when
predicting if $k$-unobserved stocks are traded vs. $k$. The curves
show the results using network features, price change features, and
all features combined.
\label{NumDaysConsecutiveTrades_vs_accuracy_features_network_pricechange}}
\end{figure}

\section{Conclusion}

Network science has examined the reaction of networks to internal stresses, particularly nodal loss, but has given considerably less attention to the relationship between external shocks in a network of stable members \cite{Saavedra08}.  Using all the instant messages among stock traders in an investing organization and their outside contacts to define structural, cognitive, and affective properties of their social network, we found that shocks --- in the form of extreme price changes --- were not associated with conventional adaptive network responses to uncertainty.  Rather, the network turtled up.  Relationships within the network favored strong ties, high clustering, and company insiders.

Implications of this work relate to networks facing disruptive environments and ``normal accidents" \cite{Perrow}.  One often cited benefit of networks is that they are more agile than hierarchies and more coordinated than markets in solving collective action problems \cite{Uzzi, Powell}.  While one basis for this benefit has been to show that institutions organized as networks do better than hierarchies in turbulent environments, there has been little work on the actual network dynamics that arise in organizations facing environmental disruptions.  Indeed, the 2013 DARPA ``robolympics" challenge was instituted to investigate whether machines can work hand in hand with humans to address problems of organizing in the face of ``normal accidents" better than humans can on their own.  Our case indicated that networks facing shocks turtle up rather than open up.  Nevertheless, we find that networks are relatively elastic --- turtled up states return to normal states relatively fast.  Whether the combination of these changes leads to better or worse performance relative to a set criterion beyond the metrics we investigated, and provided by theory, is a logical research extension and a broad goal for improving knowledge about the scientific functionality and practical management of social networks.  

Moreover, we find that the network structure 
is diagnostic of important patterns of behavior ---
including the emotional and cognitive content of individual communications, local optimality of transactions, and the sudden execution of new transactions. 
It is noteworthy that the structure of the networks 
is more effective at predicting these
behavioral patterns than the price changes in the market. 
This suggests that the network-level changes we observe are not simple
offshoots of the underlying price changes, but instead that they carry
additional rich information that can be used to analyze
the organization's behavior.
Understanding the network's reaction to shocks can thus be an
important factor in understanding the organization more broadly.

The role of the network in this analysis raises a number of 
further open questions. In particular, while there is a clear
relationship between changes in the market and changes in the network,
it is interesting to consider what the lower-level mechanisms that produce these effects might
be.  As we come to better understand
the links between shocks, networks, and behavior, we can thus arrive
at a clearer picture of networks in the context of their
surrounding environments.

\section{Acknowledgements}
This research was sponsored by the Northwestern University Institute on Complex Systems (NICO), the U. S. Army Research Laboratory and the U. S. Army Research Office under grant number W911NF-09-2-0053, Defense Advanced Research Projects Agency grant BAA-11-64, a Simons Investigator Award, a Google Research Grant, a Facebook Faculty Research Grant, an ARO MURI grant "QUANTA: Quantitative Network-based Models of Adaptive Team Behavior", and NSF grant IIS-0910664. The views and conclusions contained in this document are those of the authors and should not be interpreted as representing the official policies, either expressed or implied, of the Army Research Laboratory or the U.S. government. All our summary statistics and programs are available on request. Northwestern University IRB Approved the Study (\#STU00200578).  All data were previously collected, accessed from the firm's archive, anonymized, and involved no manipulation or interaction with subjects.  Subjects knew the data were collected and available for research purposes.  The firm provided the data under written agreement for research purposes contingent on firm's identifying characteristics remaining confidential and anonymous.

\balancecolumns

\end{document}